\documentclass[journal]{IEEEtran}
\usepackage{graphicx}
\usepackage{tabularx}
\usepackage{cite}
\usepackage{amsmath}
\usepackage{amssymb}
\usepackage{textcomp}
\usepackage{enumitem}
\usepackage{booktabs}
\usepackage{multirow}
\usepackage{siunitx}
\usepackage{balance}
\usepackage{xcolor}
\usepackage[para,online,flushleft]{threeparttable}

\DeclareMathOperator{\E}{\mathbb{E}}

\ifCLASSINFOpdf
\else
\fi

\begin{document}

\title{An Ensemble Learning Approach for In-situ Monitoring of FPGA Dynamic Power}

\author{Zhe~Lin,~\IEEEmembership{Student Member,~IEEE,}
        Sharad~Sinha,~\IEEEmembership{Member,~IEEE,}
        and~Wei~Zhang,~\IEEEmembership{Member,~IEEE}\vspace{-8mm}
\thanks{The authors Zhe Lin (zlinaf@ust.hk) and Wei Zhang (wei.zhang@ust.hk) are with the Dept. of Electronic and Computer Engineering, Hong Kong University of Science and Technology (HKUST).}
\thanks{The author Sharad Sinha (sharad\_sinha@ieee.org) is with the Dept. of Computer Science and Engineering, Indian Institute of Technology (IIT) Goa. This work was done when he was with NTU, Singapore.}}
\maketitle

\begin{abstract}
As field-programmable gate arrays become prevalent in critical application domains, their power consumption is of high concern. In this paper, we present and evaluate a power monitoring scheme capable of accurately estimating the runtime dynamic power of FPGAs in a fine-grained timescale, in order to support emerging power management techniques. In particular, we describe a novel and specialized ensemble model which can be decomposed into multiple customized decision-tree-based base learners. To aid in model synthesis, a generic computer-aided design flow is proposed to generate samples, select features, tune hyperparameters and train the ensemble estimator. Besides this, a hardware realization of the trained ensemble estimator is presented for on-chip real-time power estimation. In the experiments, we first show that a single decision tree model can achieve prediction error within 4.51\% of a commercial gate-level power estimation tool, which is 2.41--6.07$\times$ lower than provided by the commonly used linear model. More importantly, we study the extra gains in inference accuracy using the proposed ensemble model. Experimental results reveal that the ensemble monitoring method can further improve the accuracy of power predictions to within a maximum error of 1.90\%. Moreover, the lookup table (LUT) overhead of the ensemble monitoring hardware employing up to 64 base learners is within 1.22\% of the target FPGA, indicating its light-weight and scalable characteristics.
\end{abstract}

\begin{IEEEkeywords}
Field-programmable gate array (FPGA), dynamic power modeling, in-situ monitoring hardware, feature selection, k-means clustering, decision tree, ensemble learning.
\end{IEEEkeywords}

\IEEEpeerreviewmaketitle

\vspace{-3mm}
\section{Introduction}
\label{sec:intro}
\IEEEPARstart{F}{ield-programmable} gate arrays (FPGAs) are gaining popularity in wide-ranging domains such as cloud services and data centers, where they serve as hardware accelerators for computation-centric tasks. In general, FPGAs offer great benefits compared with traditional application-specific integrated circuits (ASICs), owing to their excellent programmability and short time to market. However, it is observed that applications implemented on FPGAs give rise to a 7--14$\times$ dynamic power overhead compared with the equivalent ASIC implementation~\cite{kuon07}. As the architecture complexity and integration density of modern FPGAs continue to grow, the importance of power efficiency increases and FPGA power consumption is turning out to be a key design constraint. To overcome this issue, FPGA power management techniques, which demonstrate huge potential for saving power under a stringent power budget, are receiving considerable interest.

Static power saving techniques, including power gating~\cite{tuan06}, clock gating~\cite{huda09}, dual supply voltage~\cite{dual06}, power-aware placement and routing algorithms~\cite{hoo13}, have been proposed since an early stage. More recently, runtime power management strategies have gained a surge of interest. Dynamic voltage frequency scaling (DVFS)~\cite{nunez16} and task scheduling in FPGA-CPU systems~\cite{losch16}, have been reported as successful ways to reduce the runtime power consumption on FPGAs under a tight power constraint. These methods require full awareness of the power consumption in real-time.

An existing method to monitor runtime power consumption is to utilize dedicated power measurement devices. Although these devices can be put into use in modern FPGA systems, they suffer from some disadvantages that limit their scope of applicability. A deficiency is the requirement of additional board areas for the integration of these devices, and a more critical drawback comes with their long power detection period, usually in the order of milliseconds, which exposes their inability to support power detection in a fine-grained timescale (i.e., within hundreds of clock cycles). As a result, power detection relying on on-board measuring devices is only able to support coarse-grained power management.

Emerging technologies in integrated circuit design have motivated the realization of on-chip regulators with a short voltage scaling time, in the order of sub-nanoseconds~\cite{kim12}. With the use of on-chip regulators, many novel power management strategies at fine temporal granularity, such as fine-grained DVFS~\cite{kim12,eye11,mant16} or adaptive voltage scaling~\cite{ben15}, and dynamic phase scaling for on-chip regulators~\cite{vrscale}, have been devised for contemporary computing systems. The encouraging progress in fine-grained power management techniques is prompting designers to provide power monitoring schemes within a short time frame. Therefore, it is conceivable that an accurate, fast, yet area-efficient, power detection methodology is indispensable on the roadmap towards fine-grained and efficient power management strategies in FPGA-based systems. What's more, fine-grained runtime power profiling can aid in the identification of power-critical events, which leads to more potential for power pattern analysis and power saving.

In light of the above considerations, we aim to establish an accurate, fine-grained and light-weight dynamic power monitoring scheme on FPGAs. We note that the widely deployed linear power model~\cite{lak11,najem14} is unable to adapt itself well to non-linear power behaviors of complex arithmetic units~\cite{bog00,lee15}. We therefore develop a novel non-linear decision-tree-based power model, leveraging state-of-the-art machine learning theory. In other words, we use a decision tree regression model to learn non-linear power patterns. The decision tree model demonstrates notably higher estimation accuracy compared with the traditional linear model.

In order to take one step further towards more accurate power estimation, we propose a specialized ensemble model, exploiting one customized decision tree in each of its base learners. We partition every invocation (i.e., a hardware function call) into a set of execution cycles related to different states in a finite state machine with datapath (FSMD). Thereafter, the k-means clustering algorithm is applied to partition the set of states into multiple homogeneous clusters. Training samples for different states are accordingly divided in the same manner. Following this, an individual base learner is constructed for each cluster of states using the corresponding training samples. All the base learners are finally combined into an ensemble estimator with the overall estimate being the weighted sum of the predictions from these base learners. A generic computer-aided design (CAD) flow is devised for power model synthesis, from generating samples to selecting features, clustering states, tuning hyperparameters and combining all base learners. Furthermore, we propose a light-weight and scalable realization of the trained ensemble estimator, which can be realized in an FPGA for dynamic power monitoring on the fly. We leverage the surplus FPGA resources, apart from the target applications, to implement the proposed monitoring hardware. Our proposed method is fully applicable to commercial FPGAs.

The rest of this paper is organized as follows. Section~\ref{sec:relat} gives a comprehensive discussion about modern FPGA power modeling approaches. Section~\ref{sec:back} briefly introduces the background knowledge. Section~\ref{sec:flow} and Section~\ref{sec:model} demonstrate the complete CAD flow for developing the power model. Section~\ref{sec:hard} elaborates the hardware realization of the monitoring circuits and experimental results are discussed in Section~\ref{sec:exp}. Finally, we conclude in Section~\ref{sec:conc}.

\vspace{-2mm}
\section{Related Work}
\label{sec:relat}
Recent studies about FPGA power modeling have been conducted at two different abstraction levels --- low abstraction and high abstraction. Power modeling approaches at a low abstraction level focus on synthesizing power models in terms of the granularity of interconnects, basic operators or functions. In~\cite{li03}, a macromodel for LUTs and a switch-level model for interconnects were combined to form a monolithic power model for the target FPGA at a low abstraction level. The LUT macromodel is a precharacterized lookup model for power values, which is generated by sweeping over hundreds of input vectors and finally tracking in a table the dynamic power consumption per access to LUT according to the specific input stimuli. The switch-level interconnect model is developed by fitting a switching power formula with capacitance values and the signal transition rate extracted at each node. In~\cite{naj12}, the same idea was employed, but power lookup models were generalized for arithmetic operators (e.g., adders, LUT-based multipliers and  digital signal processing (DSP)-based multipliers) with different operating frequencies, signal switching activities and data widths.

The development process of the power models at a low abstraction level is usually time-consuming. Low-level power models are also specialized in their targeted FPGA families and devices, making them difficult to migrate to other FPGA families. Moreover, low-level power models, such as those proposed in~\cite{li03} and~\cite{naj12}, usually require intensive computation to calculate power for every component independently. This issue adds to their inefficiency considering real-time detection.

In contrast, power models obtained from a high abstraction level expedite the developing time, because they can dispense with the need to dig deep into low-level power behaviors related to the devices' internal structures. Besides this, the runtime computational complexity of high-level models is much lower, widening their applicability in real-time power estimation on FPGAs. In high-level power modeling methods, signal switching activities are extracted in different time periods, and simultaneously, runtime power values are captured for the corresponding periods. Regression estimators are then trained to associate different signal activities with specific power patterns.

Most of the recent work investigating high-level power models on FPGAs~\cite{lak11,najem14,kapow16} has employed a linear regression model. In~\cite{lak11}, the IO toggle rate and resource utilization were synthesized into a linear regression model in a log-arithmetic format. This work attempted to formulate a generic power model which is apt for all applications. Nevertheless, design parameters deviating from the training applications, such as inconsistency in the IO port width, could significantly degrade the estimation accuracy. The work~\cite{najem14} leveraged the switching activities of a set of internal signals and the corresponding power values to form a linear model specifically for a single application, with the runtime computation completed by a software program running on an embedded processor in an FPGA-CPU system. Based on~\cite{najem14}, the work~\cite{kapow16} applied an online adjustment scheme and reported higher accuracy. However, this work leaned on on-board power measurement, making it almost unsuitable for fine-grained scenarios. The hardware realization of the signal monitoring in~\cite{najem14} and~\cite{kapow16} led to an LUT resource overhead of 7\% and 9\% in terms of the tested applications, respectively, as well as incurred a workload of 5\% of CPU time. Furthermore, as studied in~\cite{bog00} and~\cite{lee15}, the power behaviors of complex arithmetic units are potentially non-linear. As a consequence, the up-to-date fine-grained linear power model exhibits intrinsic restrictions on achieving high accuracy when non-linear power behaviors are discovered with the increasing training size. This is a typical problem known as \emph{underfitting} in the machine learning theory, which implies that the applied model is too simple in its characteristics to rigorously fit an intricate training set.

Within this context, we target constructing a non-linear dynamic power model to support in-situ power monitoring, providing high estimation accuracy while inducing low overheads. To the best of our knowledge, ours is the first work to introduce an ensemble power model at a high abstraction level for FPGAs. As an extension of our prior work~\cite{linfpl}, we make a substantial improvement in power estimation accuracy by devising a novel ensemble approach, wherein each of the decomposed base learners specializes in the power prediction of particular states in the FSMD.

\section{Background}
\label{sec:back}
\subsection{FPGA Power}
The power consumption of an FPGA can be decomposed into two major components: static power and dynamic power. Static power refers to the leakage power consumption when an FPGA is powered on while no circuit activities are involved. It is highly dependent on process, voltage and temperature (PVT). Dynamic power, on the other hand, is introduced by signal transitions which dissipate power by repeatedly charging and discharging the load capacitors. Equation~(\ref{eq:dyn}) formulates dynamic power $P_{dyn}$ as
\begin{equation}
\label{eq:dyn}
P_{dyn}=\sum_{i \in I}\alpha_iC_iV_{dd}^2f,
\end{equation}
which is a function of signal switching activity $\alpha_i$, capacitance $C_i$ on the net $i$ within the whole set of nets $I$, supply voltage $V_{dd}$ and operating frequency $f$.

\vspace{-3mm}
\subsection{Decision Tree and Ensemble Learning}
\begin{figure}[t]
\begin{center}
\includegraphics[width=\linewidth]{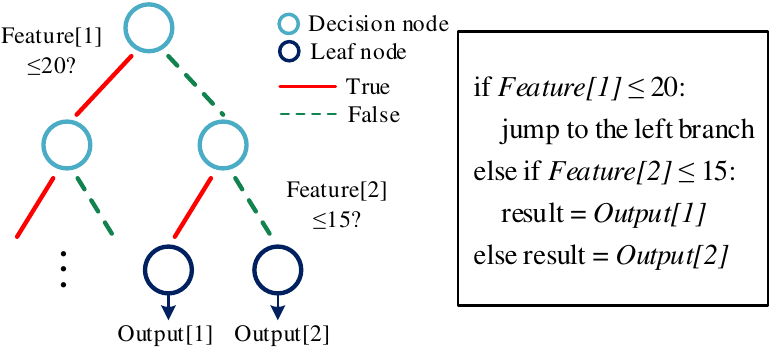}
\vspace{-6mm}
\caption{Graphical and textual representation of a decision tree.}
\label{fig:dtback}
\vspace{-6.5mm}
\end{center}
\end{figure}

The decision tree is a nonparametric hierarchical model in supervised learning for both classification and regression. It learns the samples in the form of a tree structure. A tree node can be categorized as (1) a leaf/terminal node --- a node associated with an output result (i.e., a class label for classification or a value for regression) --- or (2) a decision/internal node --- an intermediate node to decide on one of its child nodes to go to. The training process is to determine an if-then-else decision rule for every decision node and an output value for every leaf node, as shown in Fig.~\ref{fig:dtback}. To make a new inference for an unsolved case, the decision tree firstly starts with the root node and moves to one of its child nodes. This process executes iteratively, until finally a leaf node with inference result is reached. From the aspect of hardware implementation, the inference process of a decision tree can be eventually decomposed into a series of comparisons based on decision rules, and it is intrinsically easy to be implemented in a hardware-friendly manner.

Ensemble algorithms combine the predictions of several individually trained base learners in order to improve model robustness and prediction accuracy over a single estimator. In ensemble algorithms, decision trees are widely deployed as effective base learners, e.g., in random forests~\cite{breiman01}.

\vspace{-3mm}
\subsection{K-means Clustering}
Clustering is a classic unsupervised problem in machine learning theory, with the objective to separate a finite unlabeled sample set into a number of groups according to some sort of homogeneity in the samples. K-means clustering is a popular clustering method that has been widely used in data mining and industrial applications. ``K-means clustering algorithm'' usually refers to Lloyd's algorithm~\cite{lloyd}. In k-means, the cluster number $K$ is a user-specified parameter. The algorithm seeks to determine $K$ centroids in the centroid set $C$ that minimize the within-cluster sum of squares for data points in the sample set $X$, as represented in Equation~(\ref{eq:inertia}):
\begin{equation}
\label{eq:inertia}
\sum_{x \in X}\min_{c \in C}{ \lVert \mathbf{x-c} \rVert }^2.
\end{equation}

The k-means algorithm firstly begins by arbitrarily choosing k samples as the initial centroids. Each remaining sample is then assigned to its nearest centroid. After all samples are associated with centroids, every centroid is updated as the mean value of all the samples assigned to it. These two steps, namely, sample assignment and centroid update, are repeated until all centroids converge at some specific locations.

To expedite the convergence of k-means, k-means++~\cite{kpp} is used in this paper, which improves the initialization of centroids. It selects a point as a centroid based on a probability proportional to this point's contribution in the within-cluster sum of squares. This method aims at making the centroids distant from each other, resulting in a higher probability of faster convergence than randomly selecting samples to be centroids in the conventional k-means algorithm.

\section{Computer-aided Design Flow}
\label{sec:flow}
\begin{figure*}[t]
\begin{center}
\includegraphics[width=17cm]{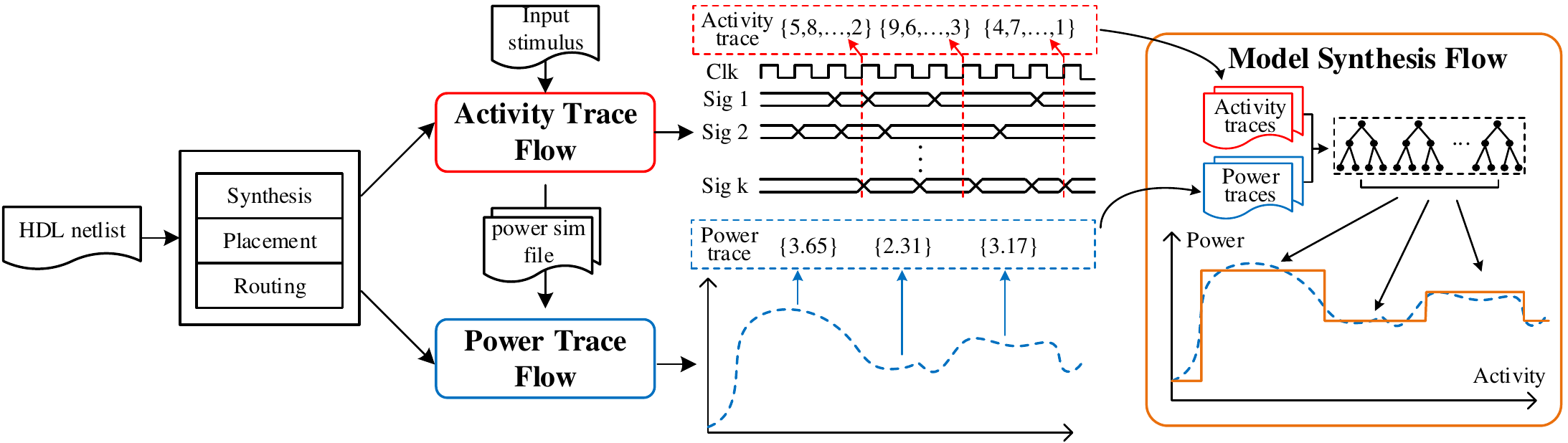}
\vspace{-3mm}
\caption{Overview of the computer-aided design flow.}
\label{fig:flowoverall}
\vspace{-7mm}
\end{center}
\end{figure*}

We propose a CAD flow to establish the power model using state-of-the-art machine learning techniques. The CAD flow is shown in Fig.~\ref{fig:flowoverall}. The design flow starts from the synthesis, placement and routing of the design. Note that in this design flow, we use the mapped hardware description language (HDL) netlist as the source, which can be exported after placement and routing. It basically consists of connections and primitives. The main advantages of using the HDL netlist are two-fold: firstly, it enables us to have access to lower level signals and further extract the indicative signals for power monitoring; and secondly, it preserves the mapping of the design, and therefore, the later integration of the monitoring hardware does not permute the original mapping.

The complete design flow can be decomposed into three sub-flows: (1) the activity trace flow (ATF), (2) the power trace flow (PTF) and (3) the model synthesis flow (MSF). We generate activity traces and power traces using the ATF and PTF, respectively. An activity trace is defined to be the set of switching activities of some monitored signals used as power pattern indicators, while a power trace provides the average power related to a particular activity trace. An activity trace and the corresponding power trace are combined as a training sample in the MSF.
\vspace{-3mm}
\subsection{Activity Trace Flow (ATF)}
\begin{figure}[t]
\begin{center}
\includegraphics[width=0.8\linewidth]{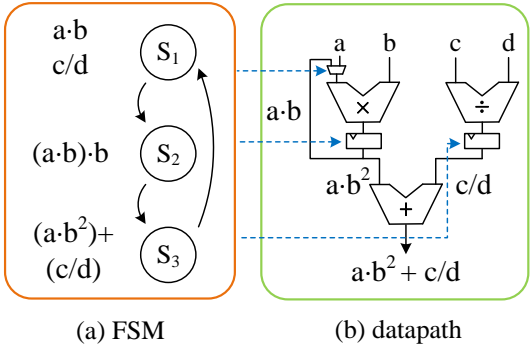}
\vspace{-3mm}
\caption{An example of FSMD implementation.}
\label{fig:fsmd}
\vspace{-7mm}
\end{center}
\end{figure}

Generally speaking, the behaviors of a digital system implemented in an FPGA can be described as an FSMD~\cite{lee15,legup,vivadohls}, as shown in Fig.~\ref{fig:fsmd}, which can be decomposed into several finite state machine (FSM) states to provide control signals for the datapath. In this paper, we limit our scope to FSMD-based hardware designs.

To build an ensemble model, we monitor the signal activities in the granularity of a state. To support this, FSM identification should be conducted to automatically extract the state register of an FSMD. We note that Odin II~\cite{odinii}, the Verilog synthesis tool in Verilog-to-Routing (VTR)~\cite{vtr}, has implemented an FSM identification algorithm by finding a combinational feedback loop from the candidate register to itself, based on a form of abstract syntax tree (AST)~\cite{ast}. The AST profiles the Verilog design and decomposes it into a hierarchical representation of different syntaxes, which can be used to identify different functional structures. We extend the basic FSM identification algorithm in Odin II to analyze feedback loops between more than one candidate register so that it is possible to idenfity FSMs written in different styles (e.g., using current\_state and next\_state registers in an FSM).

\begin{figure}[t]
\begin{center}
\includegraphics[width=0.82\linewidth]{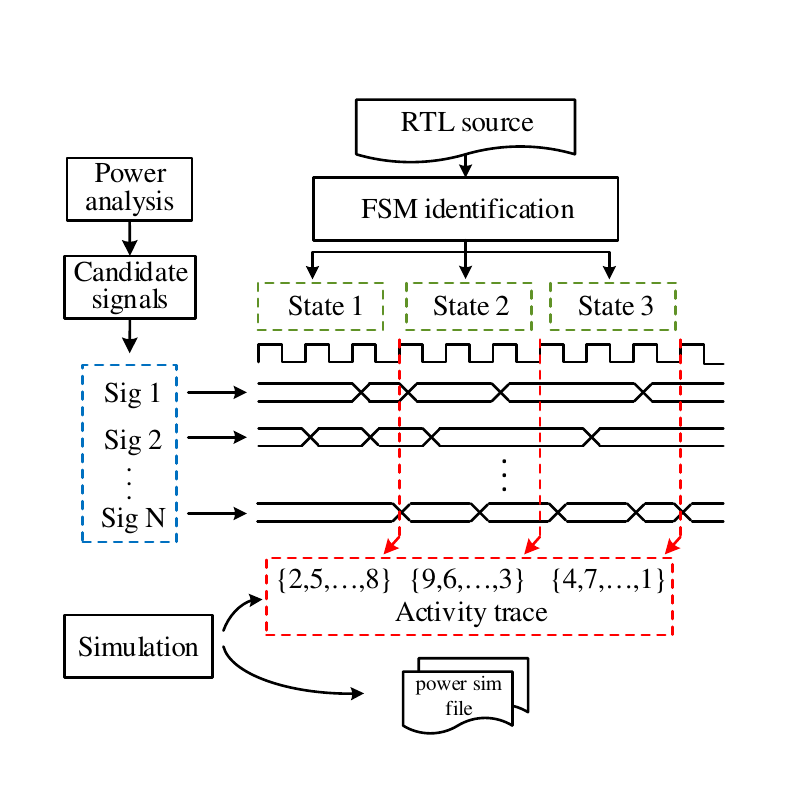}
\vspace{-3mm}
\caption{ATF overview.}
\label{fig:flowatf}
\vspace{-6mm}
\end{center}
\end{figure}

In the ATF shown in Fig.~\ref{fig:flowatf}, we aim to extract a key set of single-bit signals, out of the hundreds of internal and IO signals in the netlist, whose activities show a great influence on the dynamic power. To select the signals to monitor for general-purpose applications, we run a vector-based timing simulation with randomly generated input vectors. The simulation only stops when the simulation time is two orders of magnitude longer than the execution time of each invocation. Moreover, we incorporate a counter for each state to record its coverage. In the experiments, we have covered all states in the evaluated benchmarks. More sophisticated methods~\cite{friedman02} can be used to expedite full state traversal, which is complementary to our work and does not affect the design flow.

Signal activities are quantified and sorted through power analysis. We select from the signal list an abundance of candidate signals ($\geqslant$1,000) showing the highest switching activities to monitor, because they tend to show wide-ranging behaviors matching dynamic power patterns. In Section~\ref{subsec:emfeature}, we find that the necessary signal number is orders of magnitude smaller than the number of monitored candidate signals.

An activity trace records the activities of all the monitored signals in a sampling interval, which can be formulated as an activation function $\boldsymbol{act}(\cdot)$ in the form
\begin{equation}
\label{eq:act}
\boldsymbol{act}(\boldsymbol{s})=\boldsymbol{sw}(\boldsymbol{s},t_{end})-\boldsymbol{sw}(\boldsymbol{s},t_{start}).
\end{equation}
Therein, the $\boldsymbol{act}(\cdot)$ computes the change in the switching activities $\boldsymbol{sw}(\cdot)$ of the signal vector $\boldsymbol{s}$ over the sampling interval from $t_{start}$ to $t_{end}$. For each application, we extract the activity traces in the granularity of every state. At the same time as the activity traces are captured, we export power simulation files (\emph{.saif}) that are used for power estimation.

\vspace{-3mm}
\subsection{Power Trace Flow (PTF)}
In the PTF, we use the FPGA power estimator to perform power estimation for different time steps using power simulation files, based on the post-route design. Our main emphasis is on accurately modeling the FPGA dynamic power consumption, in that dynamic power tends to change more significantly for an applications in different time steps. Take the evaluated Xilinx Virtex-7 FPGA as an example: the static power consumption is within 2 W under the temperature margin from -40 to 80 degrees centigrade with Xilinx high-performance low-power (HPL) technology~\cite{hpl}, whereas the dynamic power can change drastically from less than 1 W to more than 20 W in terms of a wide range of applications with different characteristics and resource utilization.

\vspace{-1mm}
\subsection{Model Synthesis Flow (MSF)}
\begin{figure}[t]
\begin{center}
\includegraphics[width=0.7\linewidth]{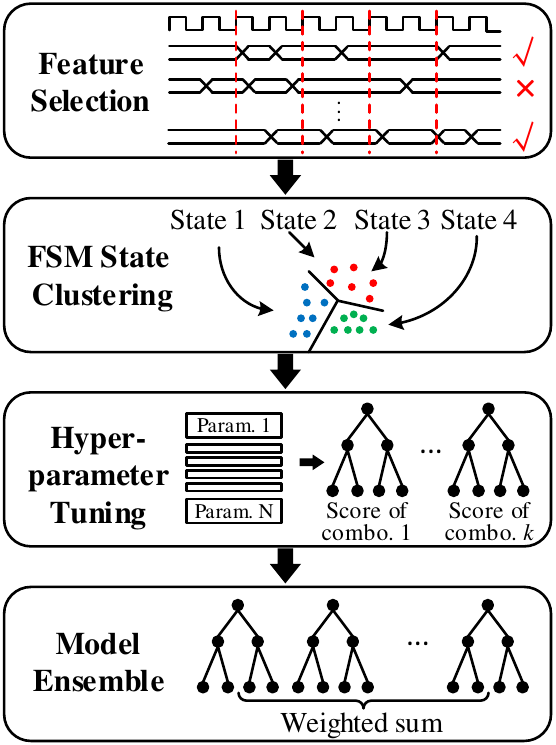}
\vspace{-3mm}
\caption{MSF overview.}
\label{fig:msf}
\vspace{-7mm}
\end{center}
\end{figure}

In the MSF, the anchor is to establish a power estimation model on the basis of the up-to-date machine learning theory, using data samples generated in the ATF and PTF. The flow for power model establishment is depicted in Fig.~\ref{fig:msf}. To begin with, feature selection is performed to diminish redundant features to monitor. Second, a k-means clustering method is applied to divide the states along with their related training samples into different clusters. Thereafter, k-fold cross validation is deployed to determine the most suitable set of hyperparameters for the decision tree estimator in every base learner. Finally, we train the base learners and combine them through a specialized ensemble approach.

\vspace{-2mm}
\section{Ensemble Model Establishment}
\label{sec:model}
This section describes the aforementioned four steps (feature selection, FSM state clustering, hyperparameter tuning and model ensemble) in the MSF for the construction of an ensemble power model. Regarding the establishment of a single decision-tree-based model, feature selection, hyperparameter tuning and decision tree training are done in the MSF, whereas FSM state clustering and model ensemble are skipped.

\vspace{-3mm}
\subsection{Feature Selection}
\label{subsec:fs}
A feature is defined as the switching activity of a monitored signal in an activity trace. Noticing that high switching features extracted from the ATF may be correlated (e.g., an input and an output of the same LUT) or show repetitive patterns (e.g., the clock signal), we leverage a \emph{recursive feature elimination} algorithm to identify the key subset of features across multiple input vectors from various invocations. Recursive feature elimination seeks to reduce the number of features recursively, and has been successfully applied to other domains~\cite{guyon02,hrav}. It requires developers to specify an estimator capable of providing attributes of feature importance (e.g., coefficients for linear model). Thus, we target the CART decision tree~\cite{cart}, which is able to provide a Gini importance value to quantify the importance of each feature.

Taking the complete feature set as input, recursive feature elimination first trains a decision tree to assess the importance of every feature and prunes the features with the lowest importance values from the generated decision tree. In this work, we set the feature reduction size as one, which means that the algorithm only filters out one feature in each iteration. After that, the remaining features are used to retrain the decision tree in order to update the Gini importance of the new feature set. The least important features are further trimmed away. These two steps work repeatedly, considering smaller and smaller sets of features in each subsequent iteration until eventually a user-specified number of critical features remains. In experiments, we study the impact of the number of selected features on the estimation accuracy of the ensemble model. Experimental results, presented in Section~\ref{subsec:emfeature}, show that around 30 key features out of more than 1,000 candidates are sufficient to achieve high accuracy.

\vspace{-3mm}
\subsection{FSM State Clustering}
During the synthesis stage, specific resources are scheduled and bound with each state in the FSMD. It is reasonable to infer that execution in different states exhibits different power consumption. This can be explained by the variation in the states' operations and utilized resources. For instance, in the datapath shown in Fig.~\ref{fig:fsmd}(b), the operations (e.g., addition, multiplication and division) invoked in various states lead to different and state-specific power consumption. Moreover, the power changes indicated by some infrequently changing but critical control signals (e.g., the start bit) are reflected in the state transitions.

Given an FSM state set containing $N$ states, $S = \{s_1, s_2,...,s_N\}$, and a user-specified cluster number $K$, we apply the k-means algorithm to partition the states into different clusters within the cluster set $C = \{c_1, c_2,...,c_K\}\ (K \leq N) $.  The relationship between the FSM state set $S$ and the cluster set $C$ can be formulated as a fuction $r(\cdot)$:
\begin{equation}
\label{eq:kmeans}
r: S \times C \rightarrow \{0,1\}.
\end{equation}
Therein, $r(s_i,c_j)=1$ means state $s_i$ is assigned to cluster $c_j$, and vice versa. All the clusters in the set $C$ possess the following attribute:
\begin{equation}
\label{eq:kmeanscon1}
c_i \cap c_j = \phi,\ \forall i \neq j.
\end{equation}
This attribute ensures that each FSM state is assigned to one and only one cluster by k-means. Provided sufficient activity patterns from features, another attribute will also be satisfied. That is
\begin{equation}
\label{eq:kmeanscon2}
c_i \neq \phi,\ \forall i \in \{1,2,...,K\}.
\end{equation}

The clusters are developed on the basis of the homogeneity of signal activities with respect to different FSM states. As a result, the FSM states in the same cluster show high similarity in signal activities, thus revealing a close resemblance of operations (some are due to resource sharing) and power characteristics. Following this observation, we create $K$ independent training sets in the same manner: the training samples corresponding to all the FSM states in the same k-means cluster will be grouped into one training set. These sample sets are non-overlapping sets constrained by Equation (\ref{eq:kmeanscon1}). In the development process of the ensemble model, we separately build one base learner for every sample set. At inference time, each base learner predicts power solely for the cluster of states it was trained with.

\vspace{-3mm}
\subsection{Hyperparameter Tuning}
We build each base learner in the form of a decision tree. The problem of learning an optimal decision tree is known to be NP-complete under several aspects of optimality. However, there are some essential hyperparameters largely determining the model accuracy. We seek to tune a set of key hyperparameters to best suit the training set in order to enhance decision tree performance. The target hyperparameter set is listed in Table~\ref{table:param}. The most influential hyperparameter is $Maximum\_depth$, which determines the number of nodes along the path from the root node down to its farthest child node. A deeper tree indicates a higher degree of model complexity, while also making it inclined to overfit on data. In contrast, a shallower tree often fails to fit in well with a complex training set. In light of this problem, the tree depth should be customized according to different characteristics of training sets in order to strike a good balance between training and testing accuracy. Likewise, the other three hyperparameters, $Minimum\_split\_sample$, $Minimum\_leaf\_sample$ and $Minimum\_leaf\_impurity$, should be coordinated with the tree depth and used to circumvent overfitting.

To automatically tailor the best hyperparameter set for the decision tree in every base learner, k-fold cross validation~\cite{kfcv} is used. K-fold cross validation performs self-testing using the training set and offers a score to evaluate the performance of an estimator with a particular set of hyperparameters. We employ ten-fold cross validation, in which the training set is equally split into ten subsets of the same size. In each iteration, one subset is selected as a validation subset, and the remaining subsets become the training subsets. The training subsets are used to construct a decision tree given a specific set of hyperparameters in Table~\ref{table:param}, while the validation set is applied to give a score for this hyperparameter set in the form of negative mean absolute error. Ten iterations are conducted with a unique validation set each time. The overall score of the evaluated hyperparameter set is the average score of the ten iterations.

We first sweep over a large hyperparameter space through cross validation and filter out the hyperparameter sets consistently leading to low scores across different training sets, as well as those offering no improvement in comparison to trees with lower complexity, namely, trees with smaller values of $Maximum\_depth$. The hyperparameter space is eventually trimmed down as follows: \{3, 4, 5, 6, 7, 8\} for $Maximum\_depth$, \{5, 10, 15, 20\} for $Minimum\_split\_sample$ and $Minimum\_leaf\_sample$, and \{0.001, 0.01, 0.02, 0.03, 0.04, 0.05\} for $Minimum\_leaf\_impurity$.

 To train each base learner, we again perform cross validation to quantify the performance of all combinations in the hyperparameter space and finally select the hyperparameter set with the highest score ranking to build a decision tree. After cross validation, we use the complete training set to train the base learner and use a test set to quantify the accuracy.
\begin{table}[t]
\centering
\small
\caption{Hyperparameters for decision tree tuning.}
\label{table:param}
\vspace{-2mm}
\begin{tabular}{p{.2\linewidth}|p{.65\linewidth}}
    \toprule
\textbf{Name}&\textbf{Description}\\
    \midrule
Maximum depth & The maximum depth that a tree can grow to. \\ \midrule
Minimum split sample & The minimum number of samples used to split a decision node.\\  \midrule
Minimum leaf sample & The minimum number of samples necessary to determine a leaf node.\\ \midrule
Minimum leaf impurity & The minimum percentage of samples giving different output at a leaf node.\\ \bottomrule
\end{tabular}
\vspace{-3mm}
\end{table}

\vspace{-3mm}
\subsection{Model Ensemble}
From the prior steps, we separately develop a decision tree in a base estimator for every cluster of FSM states generated by the k-means algorithm. Thereafter, we introduce a specialized ensemble learning method to combine different base estimators together so as to estimate the invocation-level power consumption. We take the weighted sum of the state-related power predictions to deduce the overall power, which is the same as getting the average power over different states. In principle, ensemble learning through averaging, which is known as \textit{bagging}, is expected to provide higher prediction accuracy than an individual model~\cite{bishop06,fum05}. In theory, our customized ensemble model can be proven to improve accuracy compared with an individual model.

We define the error-free target function and the prediction function for an individual invocation-level model as $h_{inv}(\cdot)$ and $y_{inv}(\cdot)$, respectively. In addition, the error-free target function and prediction function of a base estimator are defined as $h_i(\cdot)$ and $y_i(\cdot)$, respectively, where $i\in\{1,2,...,K\}$ denotes the index of the cluster/base learner. The inference result of the single invocation-level estimator and each of the base estimators can be written as
\begin{equation}
\label{eq:invpred}
y_{inv}(\mathbf{x})=h_{inv}(\mathbf{x})+\epsilon_{inv}(\mathbf{x}),
\end{equation}
\begin{equation}
\label{eq:enspred}
y_i(\mathbf{x})=h_i(\mathbf{x})+\epsilon_i(\mathbf{x}),
\end{equation}
where $\epsilon_{inv}(\cdot)$ and $\epsilon_{i}(\cdot)$ denote the corresponding error functions and $\mathbf{x}$ is the feature vector. We use the weighted sum of the error-free functions for the base estimators to get the error-free target function $h_{ens}(\cdot)$ for the ensemble model, in which the weight of a base estimator is defined as the ratio of the prediction cycles for this base estimator to the total execution cycles in an invocation. That is
\begin{equation}
\label{eq:enserrfree}
h_{ens}(\mathbf{x})=\sum_{i=1}^K\frac{t_i\cdot h_i(\mathbf{x})}{T},
\end{equation}
where $T$ is defined as the total execution cycles in an invocation, $t_i$ as the prediction cycles for each of the base estimators $i$ and $K$ as the number of clusters preset for the k-means algorithm, which also represents the number of base estimators. Regarding the single invocation model, the sum-of-squares error can be given by
\begin{equation}
\label{eq:inverror}
E_{inv}=\E_{\mathbf{x}}[\{y_{inv}(\mathbf{x})-h_{inv}(\mathbf{x})\}^2]=\E_{\mathbf{x}}[\epsilon_{inv}(\mathbf{x})^2].
\end{equation}
To calculate the sum-of-squares error for the ensemble model, we assume no less than two clusters. Three constraints have to be satisfied and can be written as
\begin{equation}
\label{eq:enscond1}
2\le K \le T,
\end{equation}
\begin{equation}
\label{eq:enscond2}
1\le t_i \le T,
\end{equation}
\begin{equation}
\label{eq:enscond3}
\sum_{i=1}^Kt_i=T.
\end{equation}
We calculate the weighted sum of the predictions from the base estimators as the overall estimation for the ensemble model. The sum-of-squares error of the ensemble model can thus be formulated as
\begin{equation}
\begin{aligned}
\label{eq:enserror}
E_{ens}&=\E_{\mathbf{x}}[\{\sum_{i=1}^K\frac{t_i\cdot y_i(\mathbf{x})}{T}-h_{ens}(\mathbf{x})\}^2]\\
&=\E_{\mathbf{x}}[\{\sum_{i=1}^K\frac{t_i\cdot h_i(\mathbf{x})+t_i\cdot \epsilon_i(\mathbf{x})}{T}-\sum_{i=1}^K\frac{t_i\cdot h_i(\mathbf{x})}{T}\}^2]\\
&=\E_{\mathbf{x}}[\{\sum_{i=1}^K\frac{t_i\cdot \epsilon_i(\mathbf{x})}{T}\}^2].
\end{aligned}
\end{equation}
For the sake of simplicity, we assume the error of different base estimators is uncorrelated and has zero mean~\cite{lee15,bishop06}. In addition, the sum-of-squares error for each of the estimators is identical~\cite{lee15}. These assumptions can be formulated as
\begin{equation}
\label{eq:assume1}
\E_{\mathbf{x}}[\epsilon_{i}(\mathbf{x})]=0, \ \ \forall i\in\{1,2,...,K\},
\end{equation}
\begin{equation}
\label{eq:assume2}
\E_{\mathbf{x}}[\epsilon_{i}(\mathbf{x})\cdot \epsilon_{j}(\mathbf{x})]=0, \ \ \forall i\neq j,
\end{equation}
\begin{equation}
\label{eq:assume3}
\E_{\mathbf{x}}[\epsilon_{i}(\mathbf{x})^2]=\E_{\mathbf{x}}[\epsilon_{inv}(\mathbf{x})^2]=\E_{\mathbf{x}}[\epsilon(\mathbf{x})^2], \ \ \forall i\in\{1,2,...,K\},
\end{equation}
where $\epsilon(\cdot)$ defines a generic error term. With the three assumptions given above, the sum-of-squares error of the ensemble model can be simplified as
\begin{equation}
\label{eq:enserrorsimplified}
E_{ens}=\frac{\sum\limits_{i=1}^Kt_i^2}{T^2}\cdot \E_{\mathbf{x}}[\epsilon(\mathbf{x})^2]=\frac{\sum\limits_{i=1}^Kt_i^2}{T^2}\cdot E_{inv}.
\end{equation}
Taking the conditions (\ref{eq:enscond1}), (\ref{eq:enscond2}) and (\ref{eq:enscond3}) into account, we apply \emph{Root-Mean Square-Arithmetic Mean Inequality} to deduce the range of the constant term of $E_{ens}$, which is given as
\begin{equation}
\label{eq:enserrorrange}
\frac{\sum\limits_{i=1}^Kt_i^2}{T^2}=\frac{\sum\limits_{i=1}^Kt_i^2}{(\sum\limits_{i=1}^Kt_i)^2} \in [\frac{1}{K},1).
\end{equation}
The left part in (\ref{eq:enserrorrange}) is minimized when $t_1, t_2, ..., t_K$ have the equal value $\frac{T}{K}$. Under this context, the relationship between $E_{ens}$ and $E_{inv}$ can be represented as
\begin{equation}
\label{eq:ensvsinv}
E_{ens}=\frac{1}{K}E_{inv}.
\end{equation}

Equation (\ref{eq:ensvsinv}) presents the theoretical proof of the power modeling problem that determines the lowest possible error using an ensemble model, with a single invocation-level model as the baseline. It demonstrates that our proposed ensemble model can intrinsically improve the average error by a factor of $K$. Nevertheless, the proof is dependent on a fundamental assumption that the sum-of-squares error incurred by different estimators is uncorrelated, which is the ideal situation. From a practical standpoint, it is very difficult or almost impossible to completely eliminate model correlation, even though many researchers have been trying to tackle this issue~\cite{brown05,liu14}.

In this work, we enhance the performance of the specialized ensemble learning by promoting the error diversities of different learners from two aspects. Firstly, \emph{training data diversity} is introduced through a specialized resampling method: every base learner is trained with a unique sample set derived from a non-overlapping cluster of FSM states. Secondly, \emph{hyperparameter diversity among base learners} is taken into consideration by customizing the best-suited hyperparameter set for each of the decomposed base learners through k-fold cross validation.

\section{Monitoring Hardware}
\label{sec:hard}
\begin{figure*}[t]
\begin{center}
\includegraphics[width=17.5cm]{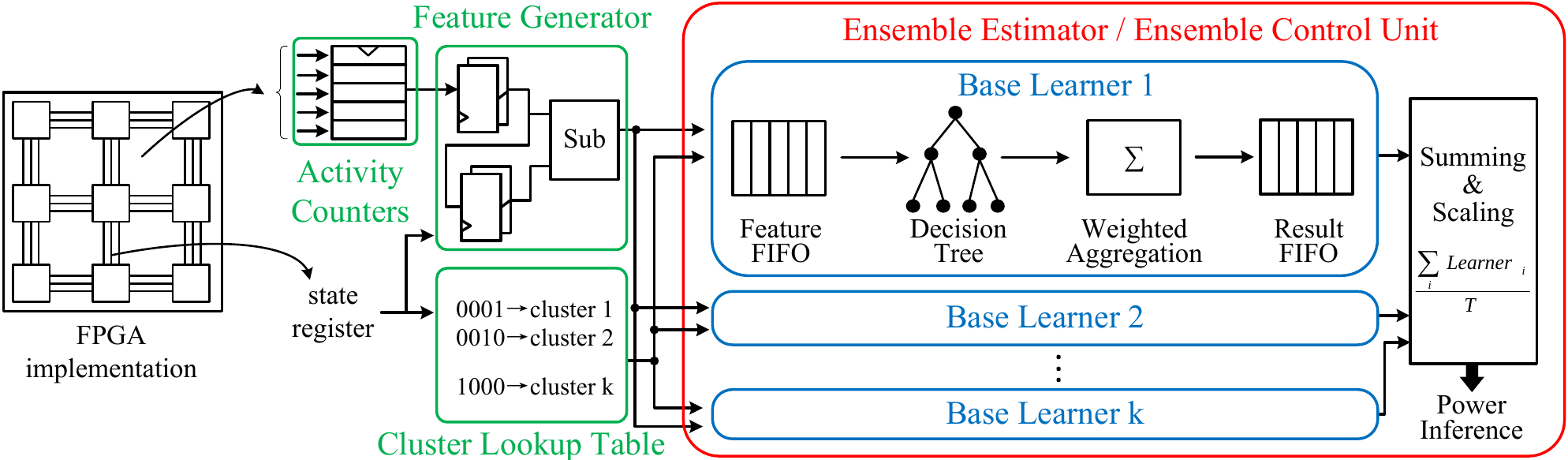}
\vspace{-2mm}
\caption{Overview of the simplified real-time monitoring hardware.}
\label{fig:enshard}
\vspace{-6mm}
\end{center}
\end{figure*}

We propose monitoring hardware to implement the trained ensemble estimator on chip and achieve dynamic power prediction on the fly. Recall that the training of the ensemble model was finished by the CAD flow discussed in Section~\ref{sec:model}. Dedicated hardware modules are devised for activity detection, state identification and power estimation at runtime. The ensemble estimator takes both the signal activities and the current state as input for power prediction, as shown in Fig.~\ref{fig:enshard}.

\vspace{-3mm}
\subsection{Preprocessing Units}
\label{subsec:pu}
A fundamental step for the ensemble hardware to function correctly is to extract activities and separate them into features related to different FSM states. To achieve this goal, three preprocessing units are designed before invoking the ensemble estimator, which are shown in the green boxes in Fig.~\ref{fig:enshard}.

\begin{figure}[t]
\begin{center}
\vspace{-1mm}
\includegraphics[width=\linewidth]{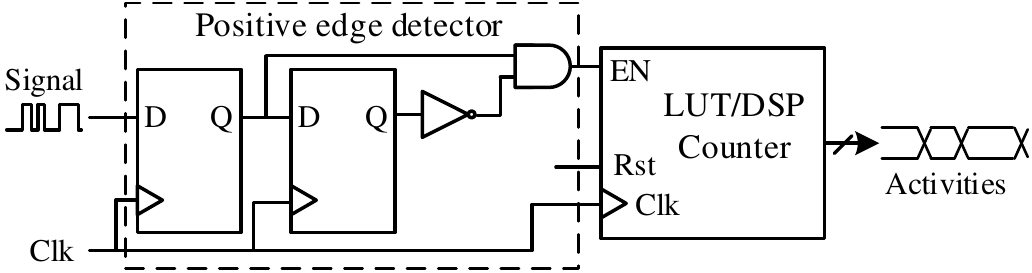}
\vspace{-7mm}
\caption{Activity counter.}
\label{fig:actcnt}
\vspace{-7mm}
\end{center}
\end{figure}

\emph{Activity counter}: Activity counters are used to capture the switching activities at runtime from the selected signals in the CAD flow. An activity counter mainly comprises a positive edge detector and a counter, as shown in Fig.~\ref{fig:actcnt}. The positive edge detector is used to detect the positive transitions of the input signal. Its output is valid for a single cycle, acting as the enable signal for the counter. We count the positive edges of the selected signals, and thereby the number of clock cycles in a sampling period is the upper bound of the signal activities, which can be used to determine the maximum bit width for the counters. We uniformly set the width as 20 bits, which can cover a wide range of sampling periods. Two types of counters are realized: an LUT counter and a DSP counter. The LUT counter is written in HDL to utilize only LUTs and flip-flops (FFs) whereas the DSP counter is instantiated using the primitive \emph{COUNTER\_LOAD\_MACRO} for Xilinx devices, a dynamic loading up counter occupying one DSP48 unit with a maximum data width of 48 bits. These two counter templates offer elasticity to developers who are aware of the application resource utilization, so that it is possible to avoid excessive use of a single type of resource.

\emph{Feature generator}: A feature generator is used to divide the switching activities into chunks specifically for each FSM state. It records the values of signal activities at the cycles when entering and exiting a state. During state transition, the feature generator performs subtraction for the two buffered activities as the way of generating features for each state.

\emph{Cluster lookup table}: The cluster lookup table stores the clustering information which is formulated as the function $r(\cdot)$ in Equation~(\ref{eq:kmeans}). Taking the state register value as input, the cluster lookup table identifies the cluster to which the executing state belongs, and generates an enable signal to correctly write the features from the feature generator to the buffer of the relevant base learner.

\vspace{-3mm}
\subsection{Ensemble Control Unit}
\label{subsec:ecu}
The ensemble control unit orchestrates different base learners for invocation-level power estimation. For each base learner, a first-in-first-out (FIFO) is used to temporarily buffer the features related to all the states within the specific cluster associated with this base learner. A decision tree regression engine, the key element in the ensemble model, implements the power prediction function $y_i(\mathbf{x}_{s_j})$, where $i\ (1\leq i\leq K)$ is the index of the cluster/base learner and $\mathbf{x}_{s_j}$ denotes the feature vector $\mathbf{x}$ for the specific state $s_j$. A summation unit is used to compute the weighted aggregation power value for a base learner. This function is defined as
\begin{equation}
\label{eq:basepow}
p(c_i) = \sum\limits_{j=1}^Nr(s_j,c_i)\cdot t(s_j)\cdot y_i(\mathbf{x}_{s_j}),
\end{equation}
where $j\ (1\leq j\leq N)$ is the index of a state, $r(\cdot)$ is the function from Equation (\ref{eq:kmeans}) and $t(s_j)$ denotes the function to extract the execution cycles in state $s_j$. Since the base learners may not operate at the same time due to the inconsistent arrival time of the input features, a result FIFO is inserted following the weighted aggregation unit in each base learner for the purpose of result sequence alignment. After all the base learners provide their predictions, the invocation-level power estimation is obtained by summing all the individual predictions and scaling down the sum by a factor of $T$ so that we get the average power among all execution cycles in an invocation. The overall power estimation is formulated as
\begin{equation}
\label{eq:invpow}
P_{ens} = \frac{1}{T}\cdot \sum\limits_{i=1}^Kp(c_i).
\end{equation}

\vspace{-3mm}
\subsection{Decision Tree Regression Engine}
\begin{figure}[t]
\begin{center}
\includegraphics[width=\linewidth]{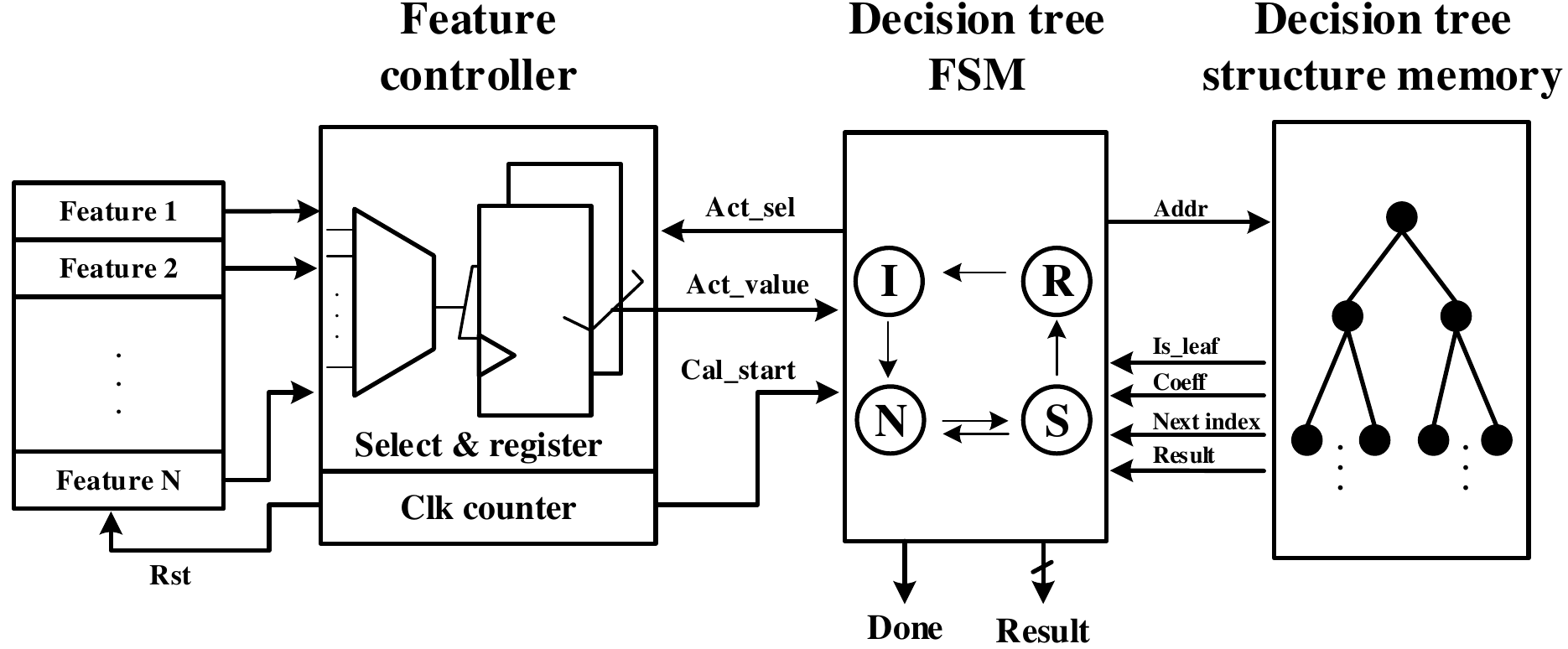}
\vspace{-7mm}
\caption{Decision tree regression engine.}
\label{fig:dtree}
\vspace{-5mm}
\end{center}
\end{figure}

The decision tree regression engine serves as the principal element in a base learner. We propose a memory-based decision tree regression engine, as shown in Fig.~\ref{fig:dtree}. There are some studies on decision tree implementation in hardware~\cite{qu14,saqib15} to maximize the throughput of decision tree computation. Our objective, however, is different from that of prior works in that the prime consideration is not throughput in our solution. Instead, we customize a decision tree structure to reduce the power and resource overheads of our monitoring hardware in the first place. The decision tree structure is completely preserved in a memory element. Additional peripheral control units are incorporated to orchestrate the feature control, tree node decoding and branch decision. To summarize, the decision tree structure in our proposed solution can be further decomposed into three subsystems: (1) a feature controller, (2) a decision tree FSM and (3) a decision tree structure memory.

\begin{figure}[t]
\begin{center}
\includegraphics[width=\linewidth]{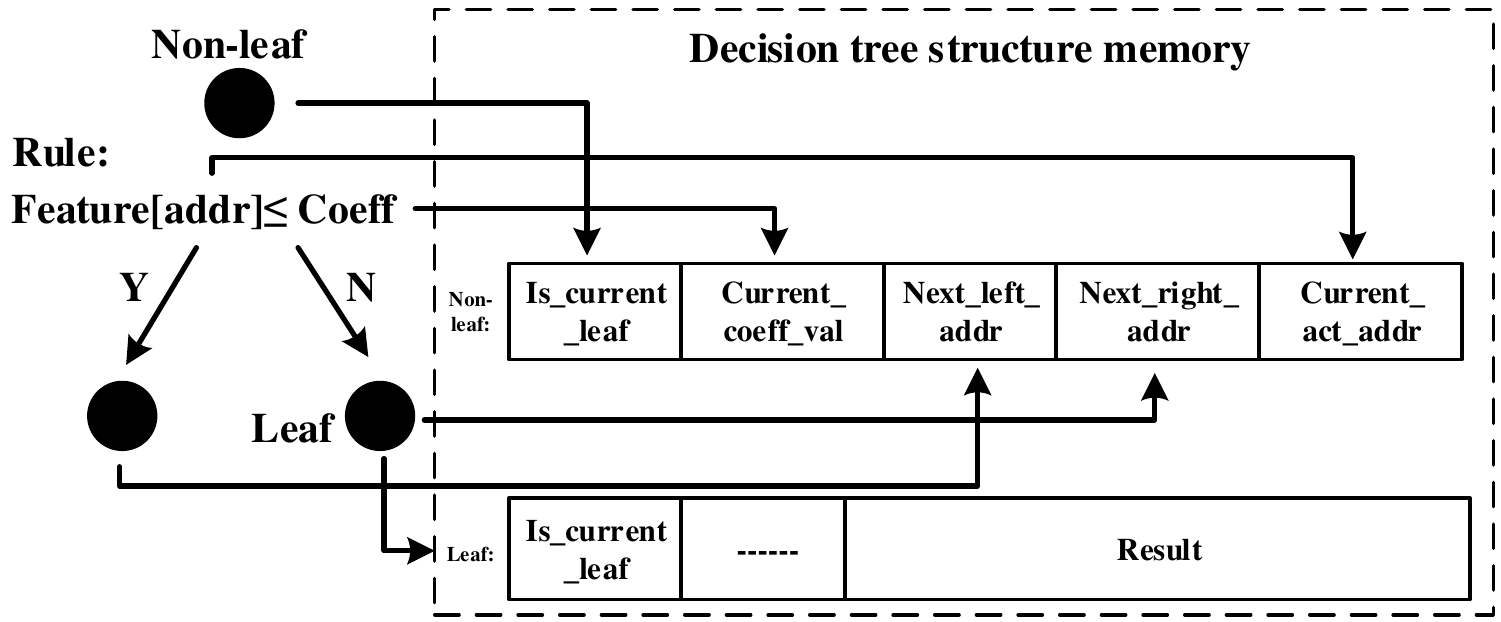}
\vspace{-6mm}
\caption{Decision tree memory structure.}
\label{fig:dtreeram}
\vspace{-7mm}
\end{center}
\end{figure}

To achieve power prediction, the feature controller buffers feature values from the feature FIFO and invokes the FSM's operating states. The decision tree FSM has four states: idle (I), node read (N), stalling (S) and result output (R). The FSM starts execution by transferring the state from idle to node read. In the node read state, the FSM fetches the node information and tree structure from the memory and completes the if-then-else branch decision by comparing the addressed feature with the decoded tree node coefficient. The stalling state will operate together with the node read state to ensure the correctness of memory reading. The execution finally terminates by switching to the result output state which sends out the estimation result and sets an indication signal high when the tree leaf has been reached. The decision tree structure memory shown in Fig.~\ref{fig:dtreeram} is the fundamental component which preserves the tree information. The memory is implemented as the block memory. For a decision node, the structure memory stores the feature address, the decision rule and child node addresses, and for a leaf node, it records the output value.

The maximum execution time for a single estimation is no more than $2n+1$ cycles, where $n$ denotes the maximum depth of the target decision tree. The decision tree structure is completely preserved in the structure memory, meaning that the proposed decision tree implementation is generally applicable to all decision tree types, varying in depths or pruning structures. Moreover, recall that a feature is the activities of a monitored signal, and therefore, it is a non-negative integer number. Correspondingly, the coefficients for comparisons in the decision rules can be represented as integers, without loss of precision. Following this observation, no floating point operations are required for the decision tree design, which contributes to the high area efficiency of its implementation, as opposed to the traditional linear model which usually requires floating point weights as well as floating point operations. Any simplification of the floating point calculation will jeopardize the precision of the linear model.

For the ensemble monitoring hardware, the overall prediction time with the largest number of base learners is 21 cycles, which can be used for power prediction under the fine granularity of several hundreds of cycles. To build a single decision-tree-based estimator, the hardware design is simplified: only activity counters and a decision tree are required. The decision tree regression engine directly fetches input features from the activity counters at the end of each sampling period and the feature controller is responsible for resetting the activity counters at the beginning of each sampling period.

\section{Experimental Results}
\label{sec:exp}

\begin{figure*}[t]
\begin{center}
\includegraphics[width=5.95cm]{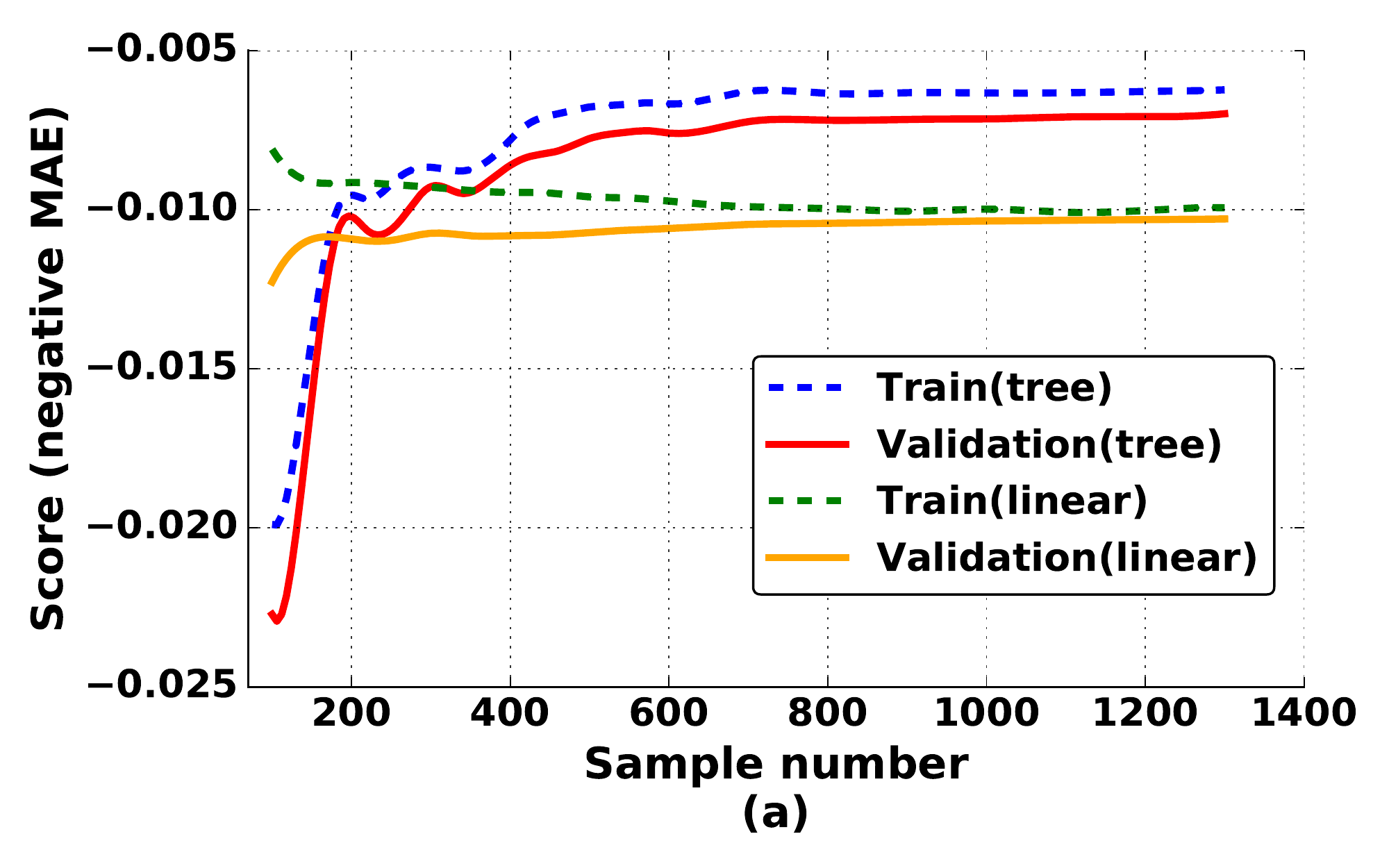}
\includegraphics[width=5.95cm]{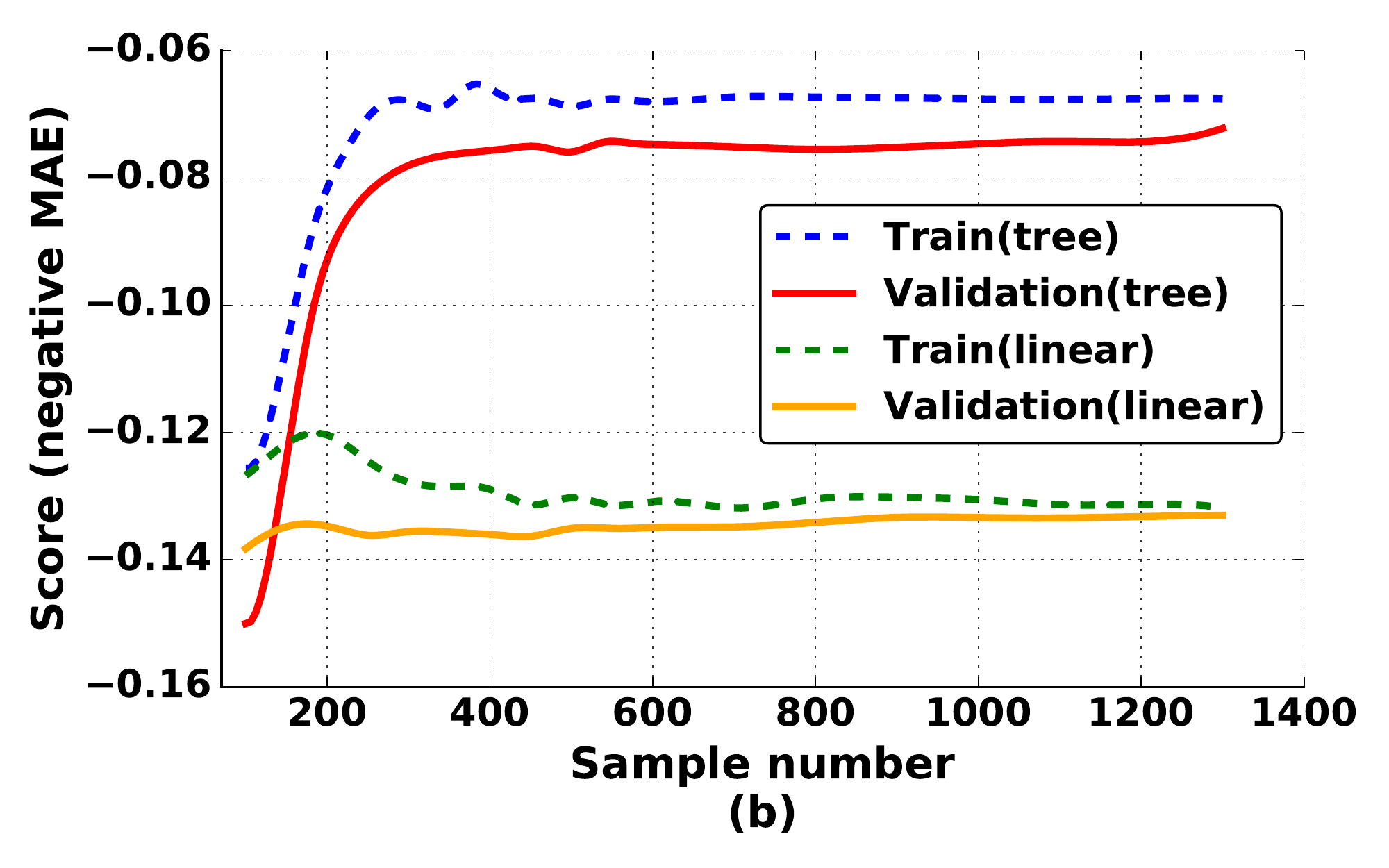}
\includegraphics[width=5.95cm]{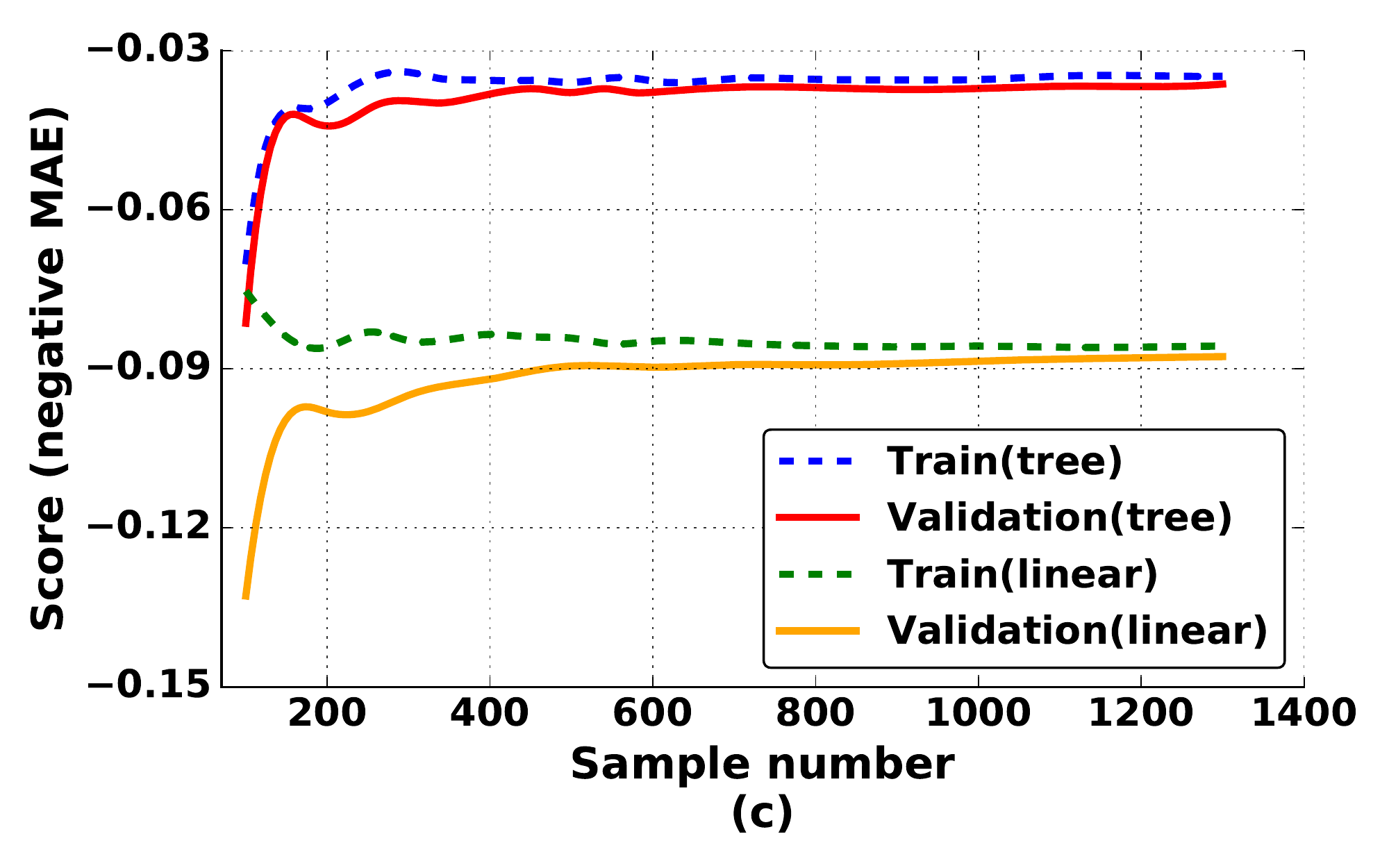}\\
\vspace{-3mm}
\caption{Learning curves of the decision tree model and linear model: (a) LUT-based; (b) DSP-based; and (c) Hybrid.}
\label{fig:learning}
\vspace{-5mm}
\end{center}
\end{figure*}

Our proposed activity trace flow is implemented in Vivado 2016.4 and the generation of the power simulation files (\emph{.saif}) is completed in Modelsim SE 10.3. The power traces are generated using Vivado power analyzer at a high confidence level~\cite{vpow}, under an ordinary temperature of 25 degrees centigrade. Note that in our experiments, we target the Xilinx tool chain and Modelsim, but the generic methodology is applicable to other vendor tools, such as Intel Quartus Prime.

The model synthesis flow is developed based on the Scikit-learn 0.18.1~\cite{scikit} machine learning toolbox. We applied our methodology to establish both the single decision-tree-based and ensemble power models, and we systematically quantify the prediction accuracy using sets of benchmarks from Polybench~\cite{polybench}, CHStone~\cite{chstone} and Machsuite~\cite{machsuite}, which are categorized by their utilized resources.

We define the following three types of benchmarks: LUT-based, DSP-based and hybrid benchmarks. The LUT-based benchmarks mainly use LUT resources, whereas the DSP-based benchmarks utilize DSPs as the major resource type. The hybrid benchmarks show a relatively balanced proportion of resource utilization of both LUTs and BRAMs or DSPs. These benchmark suites are C-based benchmarks, and we generate the synthesizable Verilog version using Vivado-HLS 2016.4. All the benchmarks are tested under a clock period of 10 ns. In the training process, we generate training samples using random distribution or normal distribution. We separate the training samples into two sets at random: 80\% of them are used as the training set and the remaining 20\% are for the purpose of testing. Customized monitoring hardware is constructed together with each benchmark on the target FPGA platform, Virtex-7 XC7V2000tflg1925-1.

In the following subsections, we first study the accuracy improvement offered by the decision tree model over the traditional linear model, and we analyze the overheads of the decision tree monitoring hardware with respect to resources, operating frequency and power. On top of that, we explore to what extent the ensemble model can further boost the estimation accuracy compared with the single decision tree model. First, we investigate the impact of feature number on power estimation accuracy. Second, we study the additional gains in accuracy as the number of base learners increases. Last, we analyze the tradeoff between overheads induced by the ensemble model and the attainable power prediction accuracy.

\vspace{-3mm}
\subsection{Decision Tree Model: Accuracy}
\label{subsec:dtaccuracy}
For the single decision tree model, we collect 2,000 samples for each benchmark. We set the sampling period to be 3 $\mu$s, which provides a good tradeoff between prediction granularity and latency (2$\times$$tree\_depth$$+$1 cycles). To achieve a fair comparison between our proposed model and the linear model~\cite{lak11,najem14}, we tailor a recursive feature elimination method for the linear model, in which the feature importance is quantified by the coefficients in the generated linear equation.

The resource utilization and the mean absolute error (MAE), in percent, for dynamic power consumption is shown in Table~\ref{table:model}. The average MAE percentage is 3.86\% and the maximum MAE percentage is 4.51\% for our proposed decision tree model, whereas those of the linear model are 14.59\% and 19.83\%, respectively. The decision tree model shows an improvement in estimation error by 2.41--6.07$\times$ in comparison to the linear model. From Table~\ref{table:model}, we can see that the improvement offered by the decision tree over the linear model is more significant for DSP-based designs, because the LUTs are essentially better fitted in with the linear regression model. In contrast, the DSPs implementing complex arithmetic operations tend to have non-linear power patterns, as reported in~\cite{bog00} and~\cite{lee15}.

\begin{table}[t]
\begin{center}
\begin{threeparttable}
\vspace{-3mm}
\caption{Resource utilization and Model accuracy.}
\label{table:model}
    \begin{tabular}[width=\linewidth]{c|c|c|c|c|c}
    \toprule
    \multirow{2}{*}{\textbf{Benchmark}} &
      \multicolumn{3}{c|}{\textbf{Resource}} &
      \multicolumn{2}{c}{\textbf{MAE} (\%)} \\
      & {LUT\tnote{1}} & \multicolumn{1}{c|}{DSP\tnote{2}} & \multicolumn{1}{c|}{BRAM\tnote{3}} & {Lin.\cite{najem14}} & {Dtree} \\
      \midrule
    \multicolumn{1}{l|}{Atax\tnote{*}\cite{polybench}} & 58691 & 0 & 0 & 12.46 & 4.14 \\
    \multicolumn{1}{l|}{Bicg\tnote{*}\cite{polybench}} & 32749 & 0 & 0 & 15.67 & 2.58 \\
    \multicolumn{1}{l|}{Bbgemm\tnote{*}\cite{machsuite}} & 261765 & 0 & 0 & 14.76 & 2.91 \\
    \multicolumn{1}{l|}{Gemver\tnote{$\dagger$}\cite{polybench}} & 9129 & 903 & 0 & 10.75 & 4.46 \\
    \multicolumn{1}{l|}{GemmNcubed\tnote{$\dagger$}\cite{machsuite}} & 10745 & 1152 & 0 & 17.80 & 4.51 \\
    \multicolumn{1}{l|}{Matrximult\tnote{$\dagger$}\cite{machsuite}} & 69125 & 2160 & 0 & 18.81 & 3.54 \\

    \multicolumn{1}{l|}{JPGizigzag\tnote{$\ddagger$}\cite{chstone}}  & 54244 & 0 & 144 & 13.33 & 4.47 \\
    \multicolumn{1}{l|}{JPGshift\tnote{$\ddagger$}\cite{chstone}}  & 9162 & 0 & 256 & 11.79 & 4.23 \\

    \multicolumn{1}{l|}{Symm\tnote{$\ddagger$}\cite{polybench}}  & 131504 & 600 & 0 & 12.06 & 4.45 \\
    \multicolumn{1}{l|}{Syr2k\tnote{$\ddagger$}\cite{polybench}} & 97932 & 800 & 0 & 19.83 & 3.58 \\
    \multicolumn{1}{l|}{Doitgen\tnote{$\ddagger$}\cite{polybench}} & 207655 & 1000 & 0 & 13.27 & 3.62 \\
    \bottomrule
  \end{tabular}
\begin{tablenotes}
\scriptsize
\item[1]Total No. LUT: 1221600\item[2]Total No. DSP: 2160\item[3]Total No. BRAM: 1292\item[*]LUT-based \item[$\dagger$]DSP-based\item[$\ddagger$]Hybrid
\end{tablenotes}
\end{threeparttable}
\end{center}
\vspace{-9mm}
\end{table}

The learning curves from cross validation further reveal the gap between the decision tree model and linear regression model regarding the capability to learn from samples. We take one from each category and show the results in Fig.~\ref{fig:learning}. Regarding the linear regression, the learning curves experience a \emph{high-bias} scenario: the error is excessively high and the linear model is unable to improve inference accuracy given more training samples. In principle, the high-bias situation means the \emph{underfitting} problem of the training data has occurred. This also accounts for the deterioration in training accuracy as non-linear power patterns continuously appear with more samples. Comparatively speaking, the decision tree model exhibits a superior ability to learn from a larger training set, due to a higher model complexity and non-linearity.

We also study the effect of operating frequency on estimation accuracy. We use the pre-trained power estimators to verify the model's adaptability as operating frequencies change. We employ the same number of estimation cycles as the prior experiment, but run our CAD flow to collect new activity and power traces under multiple frequencies and sampling periods solely for testing purposes. Noting that the model will definitely produce biased estimation under a new frequency, we calibrate the prediction by multiplying it by the ratio of the current frequency to the 100 MHz baseline frequency ($\frac{f_{current}}{f_{baseline}}$). This is based on the fact that dynamic power consumption is proportional to the operating frequency, as shown in Equation~(\ref{eq:dyn}). The degradation of error is within 0.28\% when the operating frequency varies. In all, the decision tree model offers high adaptability under a wide range of frequencies. This conclusion also holds when the decision trees are used in base learners for the ensemble model.

\begin{table*}[t]
\begin{center}
\caption{Decision tree hyperparameter settings \& Model overheads.}
\label{table:parover}
\vspace{-2mm}
\begin{tabular}[width=\linewidth]{cccccccccccccc}
    \toprule
    \multicolumn{1}{c|}{\multirow{2}{*}{\textbf{Benchmark}}} & \multicolumn{1}{c|}{\multirow{1}{*}{\textbf{Max}}} & \multicolumn{1}{c|}{\multirow{1}{*}{\textbf{Min split}}} & \multicolumn{1}{c|}{\multirow{1}{*}{\textbf{Min leaf}}} & \multicolumn{1}{c|}{\multirow{1}{*}{\textbf{Min leaf}}} & \multicolumn{1}{c|}{\multirow{1}{*}{\textbf{No.}}} & \multicolumn{4}{c|}{\multirow{1}{*}{\textbf{Resource} (in number)}} & \multicolumn{2}{c|}{\multirow{1}{*}{\textbf{Frequency} (MHz)}} & \multicolumn{1}{c}{\multirow{1}{*}{\textbf{Power}}}\\

     \multicolumn{1}{c|}{} & \multicolumn{1}{c|}{\textbf{depth}}& \multicolumn{1}{c|}{\textbf{sample}}& \multicolumn{1}{c|}{\textbf{sample}}& \multicolumn{1}{c|}{\textbf{impurity}} & \multicolumn{1}{c|}{\textbf{Feature}} & \multicolumn{1}{c|}{LUT}& \multicolumn{1}{c|}{DSP}& \multicolumn{1}{c|}{FF}& \multicolumn{1}{c|}{BRAM} & \multicolumn{1}{c|}{baseline} & \multicolumn{1}{c|}{dtree added} & \multicolumn{1}{c}{(mW)} \\ \midrule

     \multicolumn{1}{l|}{Atax}&
     \multicolumn{1}{c|}{5}&\multicolumn{1}{c|}{20}&\multicolumn{1}{c|}{20}&\multicolumn{1}{c|}{0.01}&\multicolumn{1}{c|}{7}&
     \multicolumn{1}{c|}{127}&\multicolumn{1}{c|}{7}&\multicolumn{1}{c|}{198}&\multicolumn{1}{c|}{2.5}&
     \multicolumn{1}{c|}{117.65}&\multicolumn{1}{c|}{0}& \multicolumn{1}{c}{3}\\

     \multicolumn{1}{l|}{Bicg}&
     \multicolumn{1}{c|}{5}&\multicolumn{1}{c|}{5}&\multicolumn{1}{c|}{5}&\multicolumn{1}{c|}{0.001}&\multicolumn{1}{c|}{6}&
     \multicolumn{1}{c|}{125}&\multicolumn{1}{c|}{6}&\multicolumn{1}{c|}{176}&\multicolumn{1}{c|}{0.5}&
     \multicolumn{1}{c|}{113.30}&\multicolumn{1}{c|}{-0.01}&\multicolumn{1}{c}{2}\\

     \multicolumn{1}{l|}{Bbgemm}&
     \multicolumn{1}{c|}{6}&\multicolumn{1}{c|}{5}&\multicolumn{1}{c|}{20}&\multicolumn{1}{c|}{0.001}&\multicolumn{1}{c|}{17}&
     \multicolumn{1}{c|}{187}&\multicolumn{1}{c|}{17}&\multicolumn{1}{c|}{419}&\multicolumn{1}{c|}{2}&
     \multicolumn{1}{c|}{105.62}&\multicolumn{1}{c|}{-0.38}&\multicolumn{1}{c}{9}\\

     \multicolumn{1}{l|}{Gemver}&
     \multicolumn{1}{c|}{6}&\multicolumn{1}{c|}{20}&\multicolumn{1}{c|}{5}&\multicolumn{1}{c|}{0.05}&\multicolumn{1}{c|}{12}&
     \multicolumn{1}{c|}{152}&\multicolumn{1}{c|}{12}&\multicolumn{1}{c|}{308}&\multicolumn{1}{c|}{2}&
     \multicolumn{1}{c|}{100.36}&\multicolumn{1}{c|}{+3.31}&\multicolumn{1}{c}{11}\\

     \multicolumn{1}{l|}{Gemmncubed}&
     \multicolumn{1}{c|}{5}&\multicolumn{1}{c|}{20}&\multicolumn{1}{c|}{20}&\multicolumn{1}{c|}{0.03}&\multicolumn{1}{c|}{9}&
     \multicolumn{1}{c|}{149}&\multicolumn{1}{c|}{9}&\multicolumn{1}{c|}{242}&\multicolumn{1}{c|}{2.5}&
     \multicolumn{1}{c|}{134.17}&\multicolumn{1}{c|}{+0.02}&\multicolumn{1}{c}{4}\\

     \multicolumn{1}{l|}{Matrixmult}&
     \multicolumn{1}{c|}{4}&\multicolumn{1}{c|}{5}&\multicolumn{1}{c|}{5}&\multicolumn{1}{c|}{0.03}&\multicolumn{1}{c|}{4}&
     \multicolumn{1}{c|}{108}&\multicolumn{1}{c|}{0}&\multicolumn{1}{c|}{325}&\multicolumn{1}{c|}{0.5}&
     \multicolumn{1}{c|}{133.74}&\multicolumn{1}{c|}{-3.89}&\multicolumn{1}{c}{4}\\

     \multicolumn{1}{l|}{JPGizigzag}&
     \multicolumn{1}{c|}{7}&\multicolumn{1}{c|}{5}&\multicolumn{1}{c|}{5}&
     \multicolumn{1}{c|}{0.001}&\multicolumn{1}{c|}{10}&
     \multicolumn{1}{c|}{154}&\multicolumn{1}{c|}{10}&\multicolumn{1}{c|}{265}&
     \multicolumn{1}{c|}{1}&
     \multicolumn{1}{c|}{102.60}&\multicolumn{1}{c|}{-1.32}&\multicolumn{1}{c}{8}\\

     \multicolumn{1}{l|}{JPGshift}&
     \multicolumn{1}{c|}{7}&\multicolumn{1}{c|}{5}&\multicolumn{1}{c|}{20}&
     \multicolumn{1}{c|}{0.001}&\multicolumn{1}{c|}{11}&
     \multicolumn{1}{c|}{152}&\multicolumn{1}{c|}{11}&\multicolumn{1}{c|}{286}&
     \multicolumn{1}{c|}{1}&
     \multicolumn{1}{c|}{125.08}&\multicolumn{1}{c|}{+0.17}&\multicolumn{1}{c}{11}\\

     \multicolumn{1}{l|}{Symm}&
     \multicolumn{1}{c|}{6}&\multicolumn{1}{c|}{5}&\multicolumn{1}{c|}{5}&\multicolumn{1}{c|}{0.02}&\multicolumn{1}{c|}{17}&
     \multicolumn{1}{c|}{317}&\multicolumn{1}{c|}{0}&\multicolumn{1}{c|}{1077}&\multicolumn{1}{c|}{2}&
     \multicolumn{1}{c|}{125.47}&\multicolumn{1}{c|}{+0.38}&\multicolumn{1}{c}{11}\\

     \multicolumn{1}{l|}{Syr2k}&
     \multicolumn{1}{c|}{6}&\multicolumn{1}{c|}{5}&\multicolumn{1}{c|}{5}&\multicolumn{1}{c|}{0.05}&\multicolumn{1}{c|}{20}&
     \multicolumn{1}{c|}{308}&\multicolumn{1}{c|}{0}&\multicolumn{1}{c|}{1019}&\multicolumn{1}{c|}{2}&
     \multicolumn{1}{c|}{125.02}&\multicolumn{1}{c|}{+0.17}&\multicolumn{1}{c}{9}\\

     \multicolumn{1}{l|}{Doitgen}&
     \multicolumn{1}{c|}{7}&\multicolumn{1}{c|}{5}&\multicolumn{1}{c|}{10}&\multicolumn{1}{c|}{0.01}&\multicolumn{1}{c|}{17}&
     \multicolumn{1}{c|}{222}&\multicolumn{1}{c|}{0}&\multicolumn{1}{c|}{445}&\multicolumn{1}{c|}{2}&
     \multicolumn{1}{c|}{120.48}&\multicolumn{1}{c|}{+0.55}&\multicolumn{1}{c}{11}\\
     \bottomrule
\end{tabular}
\end{center}
\vspace{-5mm}
\end{table*}

\vspace{-3mm}
\subsection{Decision Tree Model: Overhead}
We analyze the overheads of integrating the decision tree power model into each benchmark from the following three aspects: resource utilization, operating frequency and power dissipation. The decision tree hyperparameter settings and the overheads for all three types of benchmarks are presented in Table~\ref{table:parover}.

The monitoring circuits consume less than 0.05\% of LUTs, 0.2\% of block random-access memories (BRAMs) and 0.4\% of DSPs, as shown in Table~\ref{table:parover}. The decision tree exhibits extremely high area efficiency because it mainly leverages integer comparisons, while the linear model consumes a large number of floating-point additions and multiplications. The power dissipation of our single decision tree model is extremely low, namely, less than 11 mW, and can be neglected since an application generally has a power consumption in the order of watts. The maximum degradation of operating frequency is 3.89 MHz (2.91\%). There is an interesting phenomenon that in some cases, the frequencies can even be improved slightly after the monitoring hardware is added. This is attributed to the fact that the placement and routing problem is NP-complete~\cite{wu96} and the associated algorithms are pseudo-random in nature, which introduces uncertainties that may slightly improve or degrade the timing slack of the design for some corner cases, without a guarantee of global optimum. This phenomenon also appears in the ensemble model. In conclusion, the decision-tree-based monitoring hardware demonstrates low overheads in resource utilization, operating frequency and power dissipation. The monitoring hardware of a single decision tree can be efficiently integrated into register-transfer level (RTL) designs for on-chip power monitoring.

\vspace{-3mm}
\subsection{Ensemble Model: Feature Number}
\label{subsec:emfeature}
\begin{figure}[t]
\begin{center}
\includegraphics[width=0.85\linewidth]{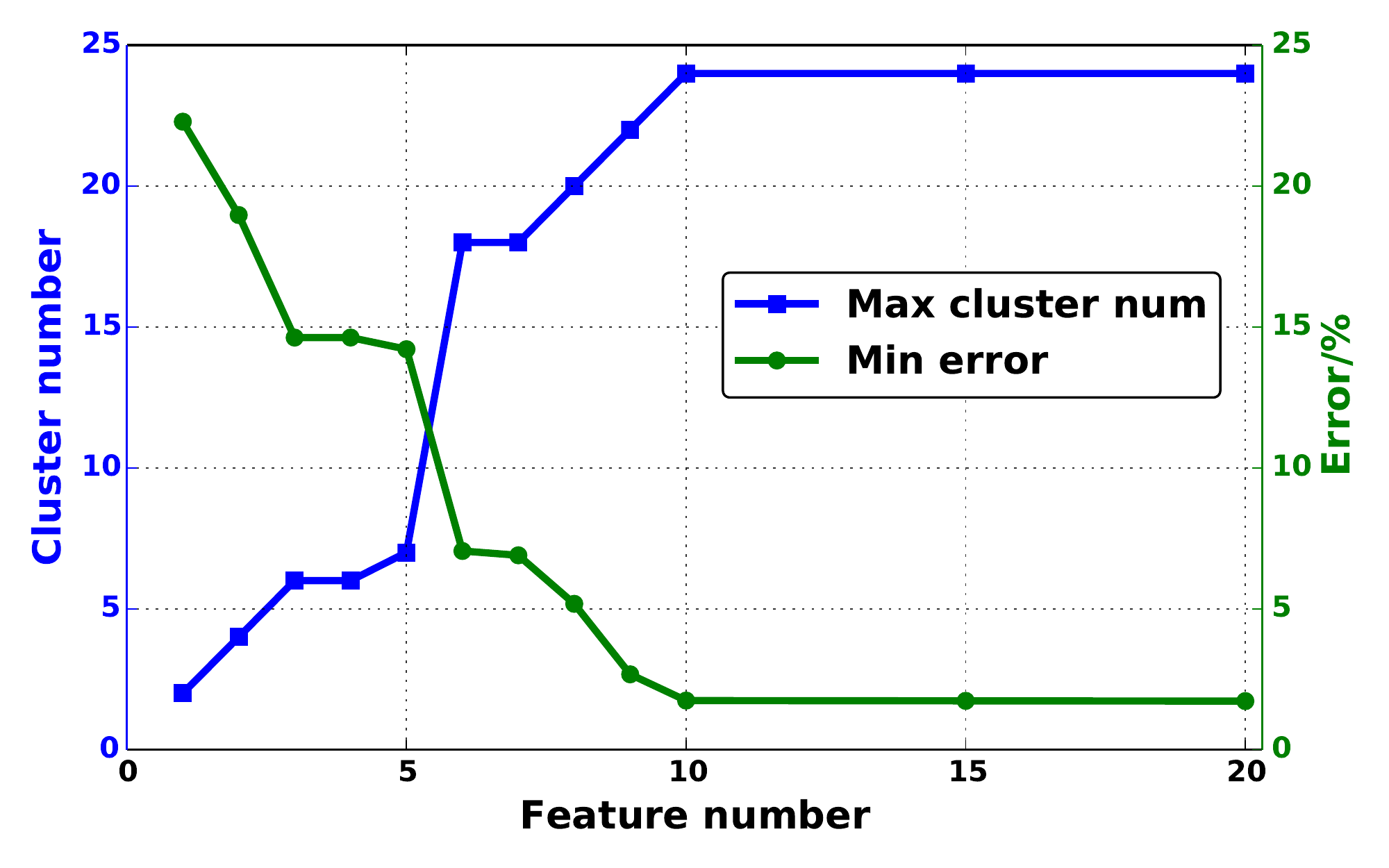}
\vspace{-3mm}
\caption{Maximum identifiable clusters and minimum estimation error with different feature numbers.}
\label{fig:feaacc}
\vspace{-5mm}
\end{center}
\end{figure}

To analyze the ensemble model, we first study the estimation accuracy when applying different numbers of features. The results for Atax are taken as an example, as shown in Fig.~\ref{fig:feaacc}, and other benchmarks exhibit a similar tendency. We observe that the maximum number of identifiable clusters from the k-means algorithm is strongly dependent on the number of features used, besides the predetermined cluster number $K$. The number of clusters from k-means can be fewer than $K$ when the feature number is small, which violates Equation~(\ref{eq:kmeanscon2}), owing to the inadequate information provided by the limited subset of features. With more activity patterns discovered through the deployment of more features, the number of maximum identifiable clusters gets larger. The minimum attainable error also reduces with the growth in identifiable clusters. Gradually, when the number of identifiable clusters becomes equal to the number of FSM states, further increase in the number of features only offers a trivial refinement in error. In light of this phenomenon, we select the number of features to be exactly enough to maximize the number of identifiable clusters. Experimental results reveal that ten features are enough for Atax, Bicg and Matrixmult, 30 features for Bbgemm and 20 features for the others. We keep these settings in the following experiments.

\vspace{-3mm}
\subsection{Ensemble Model: Accuracy}
\label{subsec:emacc}

\begin{figure*}[t]
\begin{center}
\includegraphics[width=5.95cm]{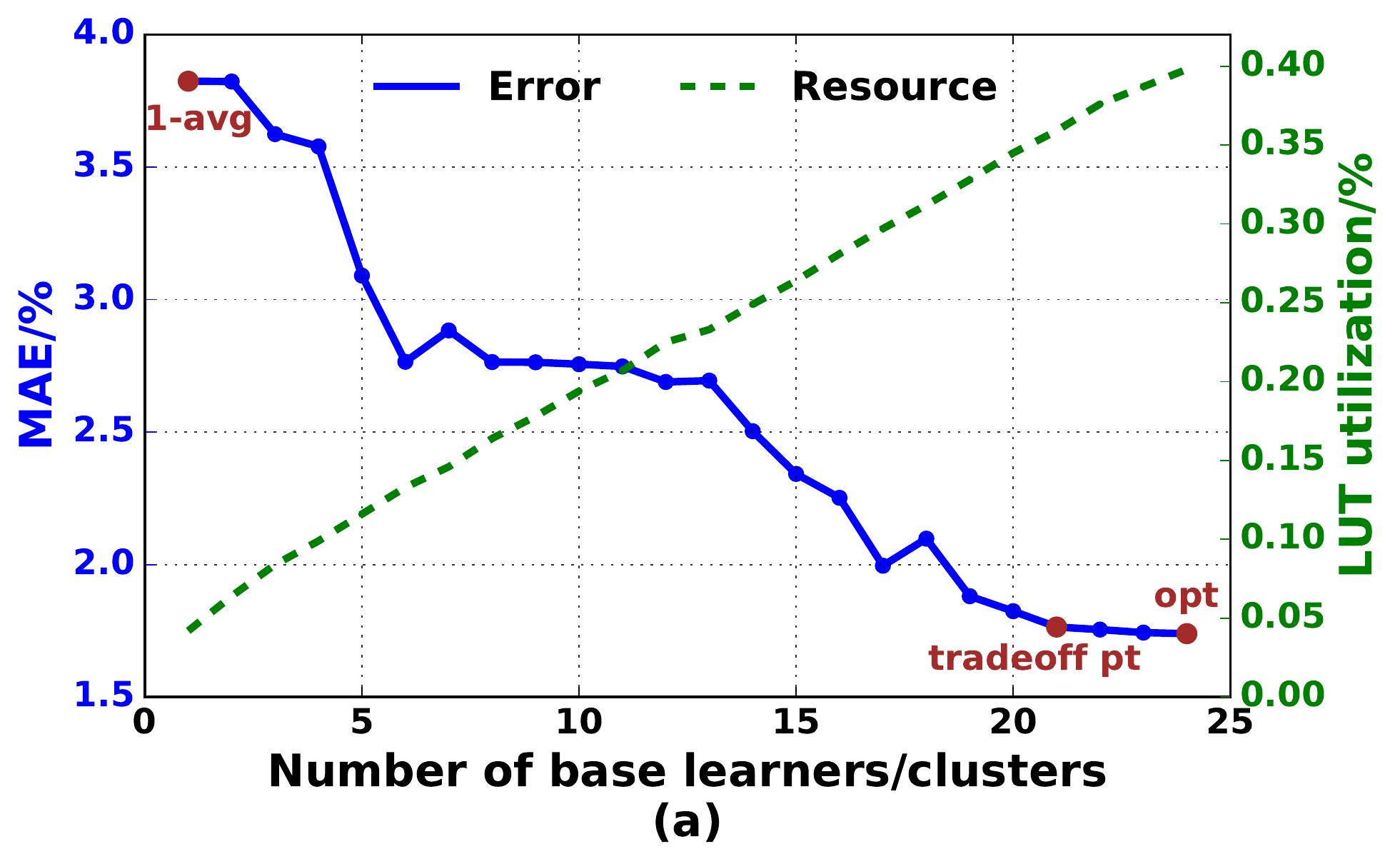}
\includegraphics[width=5.95cm]{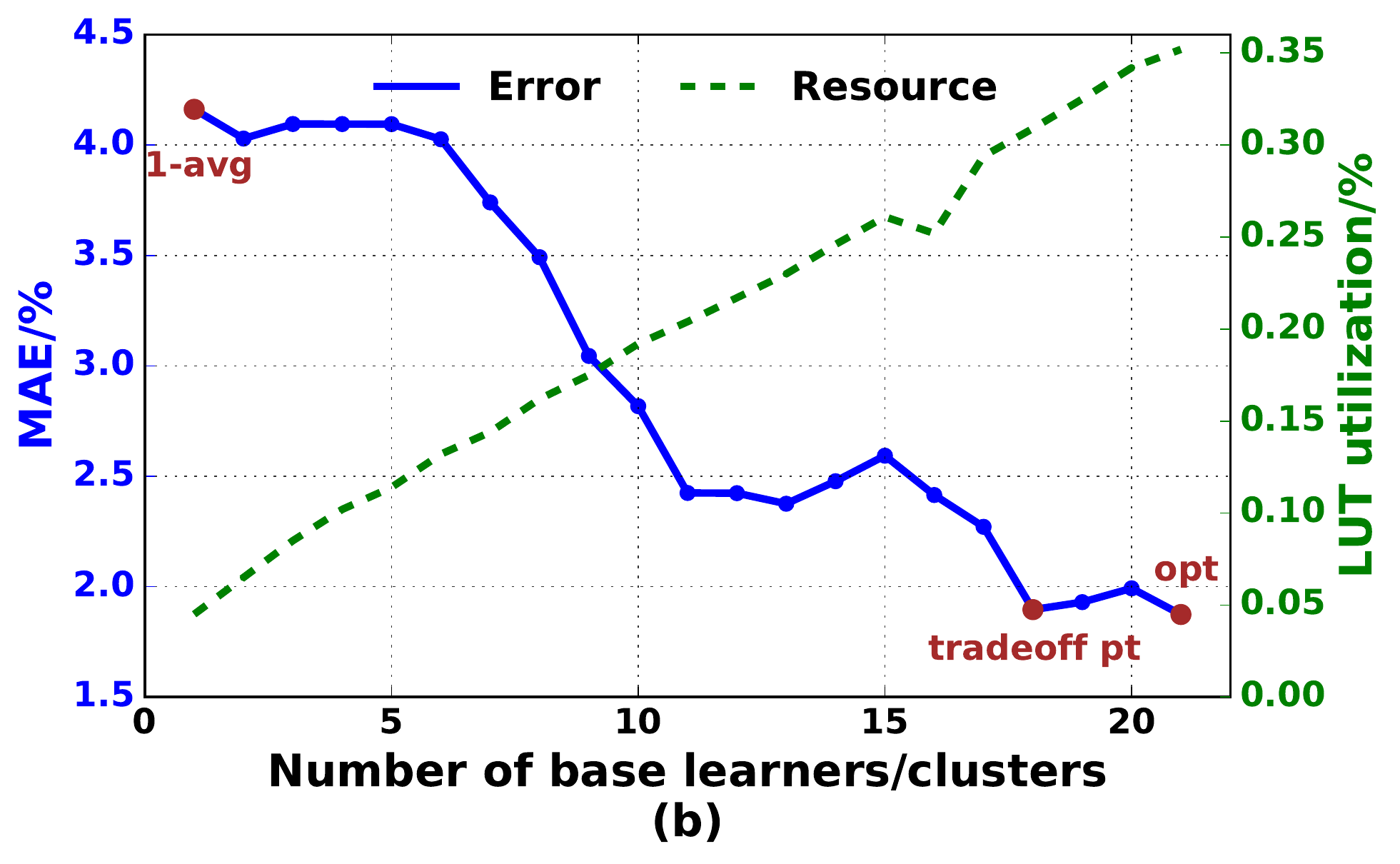}
\includegraphics[width=5.95cm]{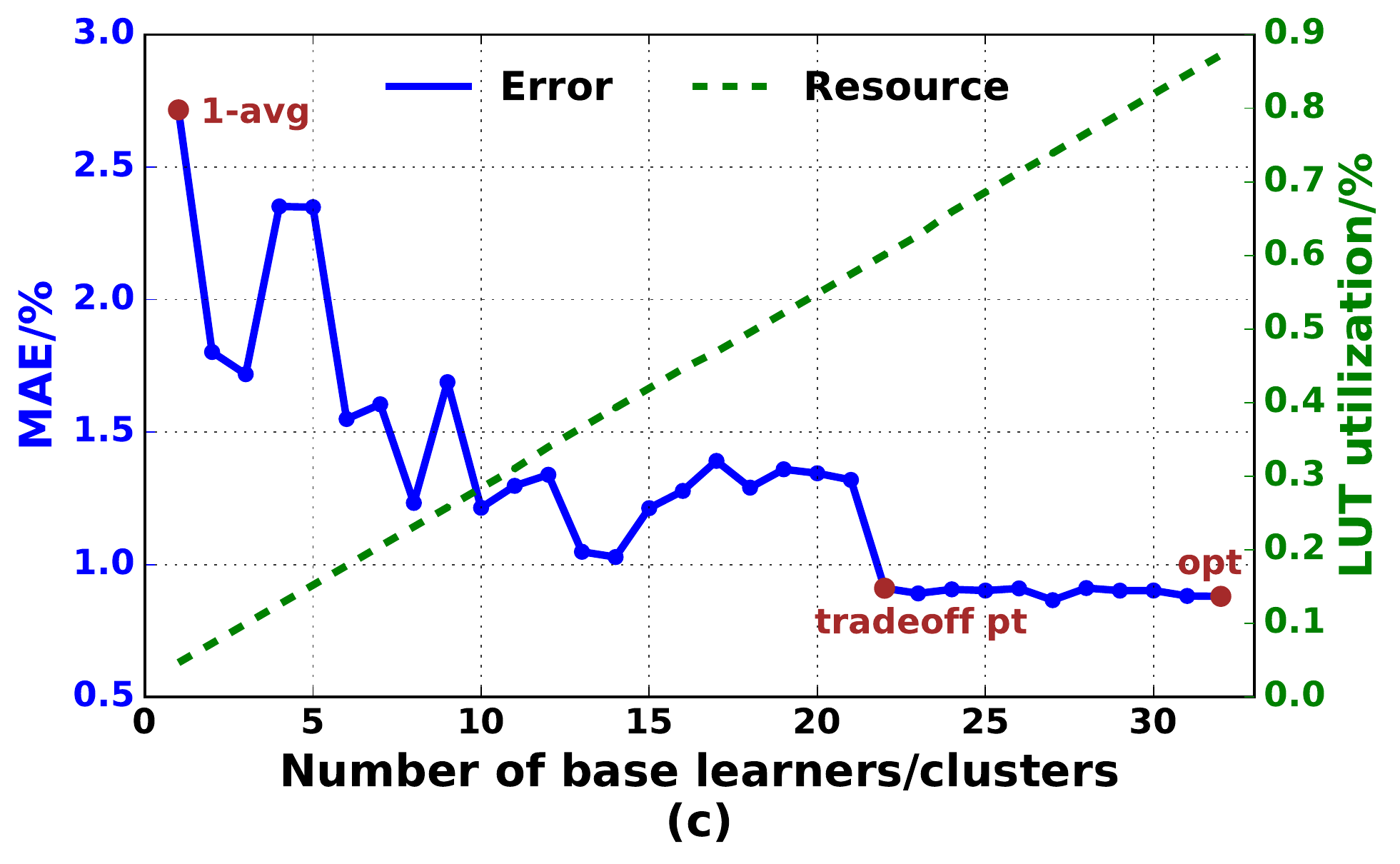}\\
\includegraphics[width=5.95cm]{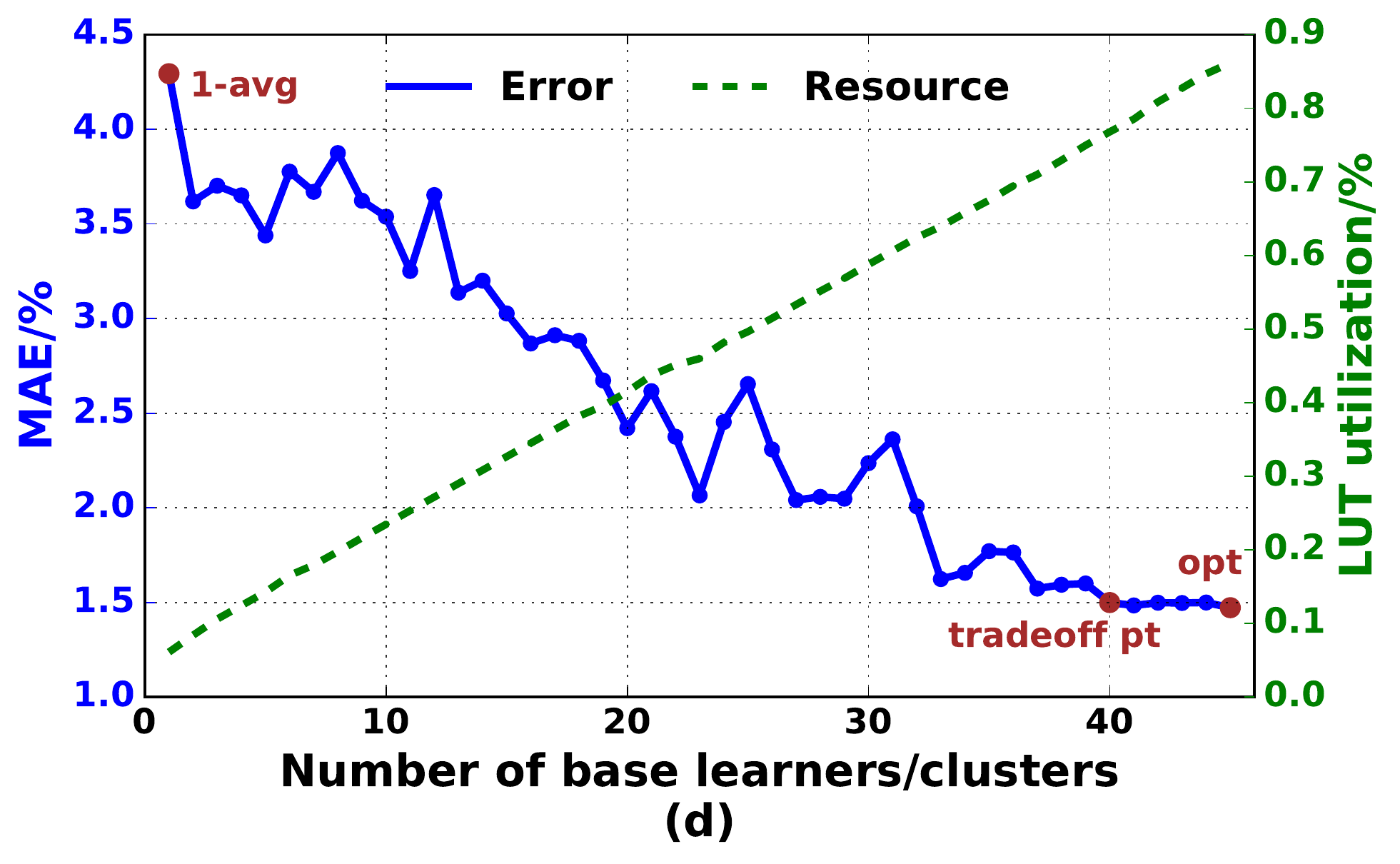}
\includegraphics[width=5.95cm]{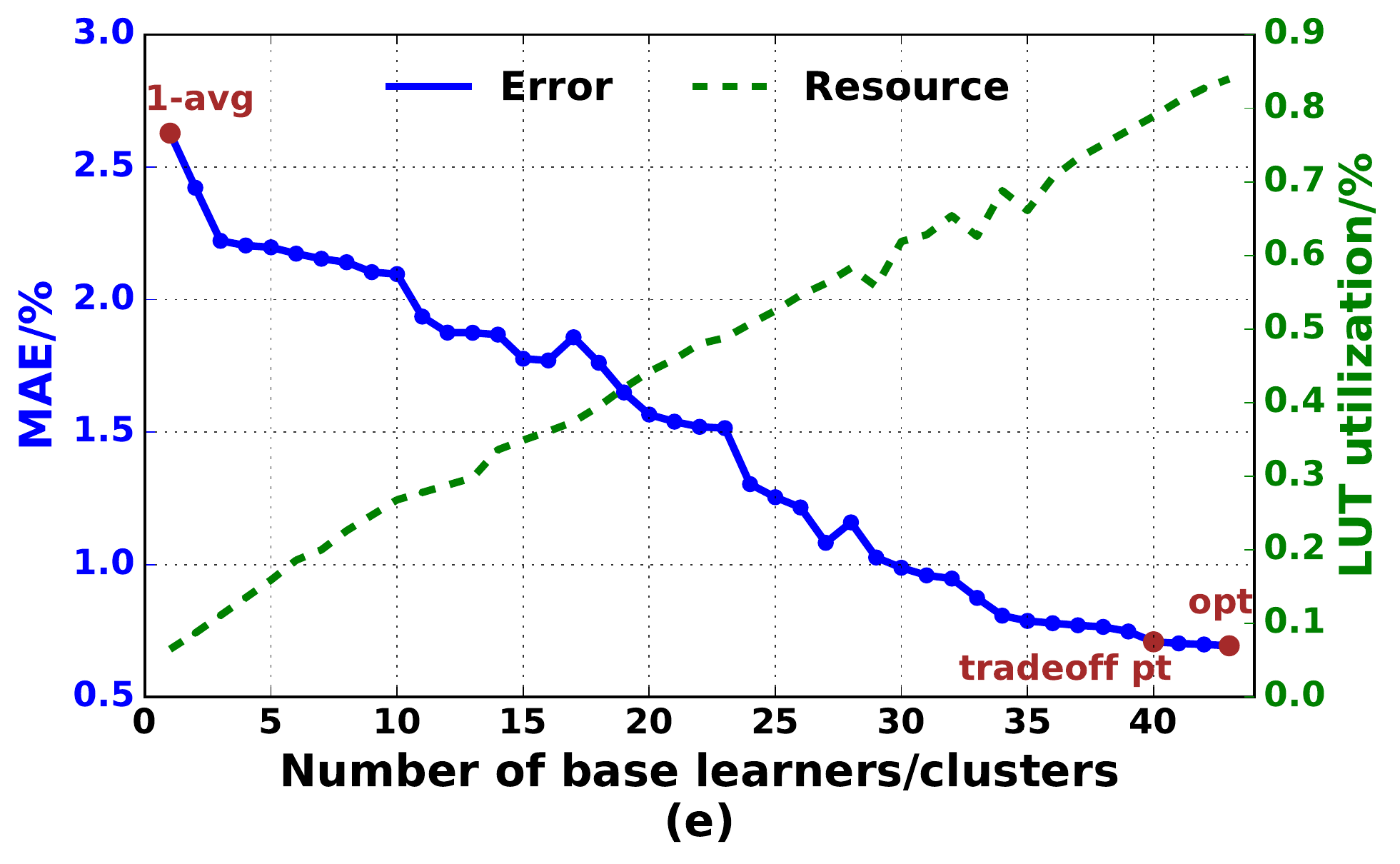}
\includegraphics[width=5.95cm]{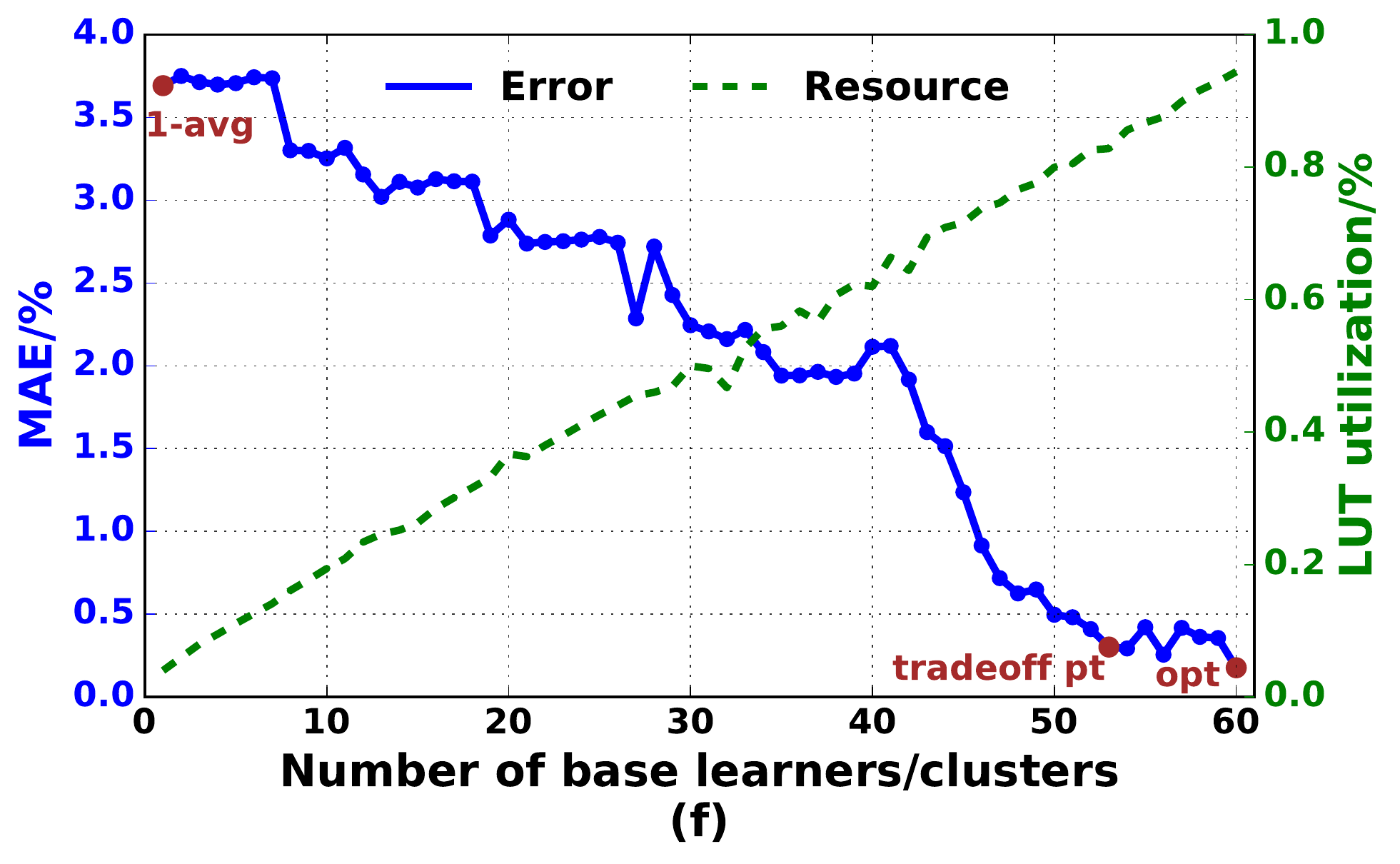}\\
\includegraphics[width=5.95cm]{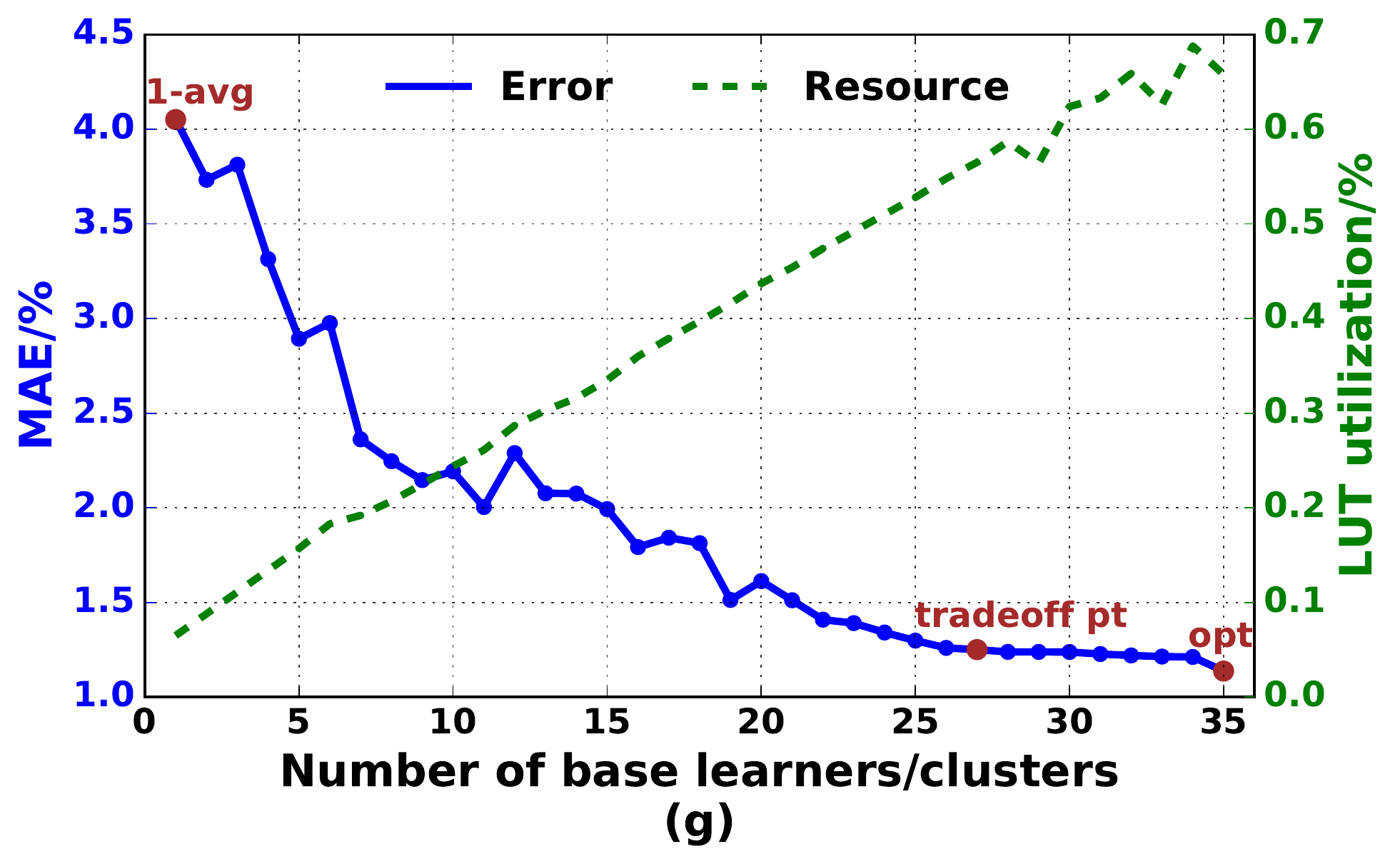}
\includegraphics[width=5.95cm]{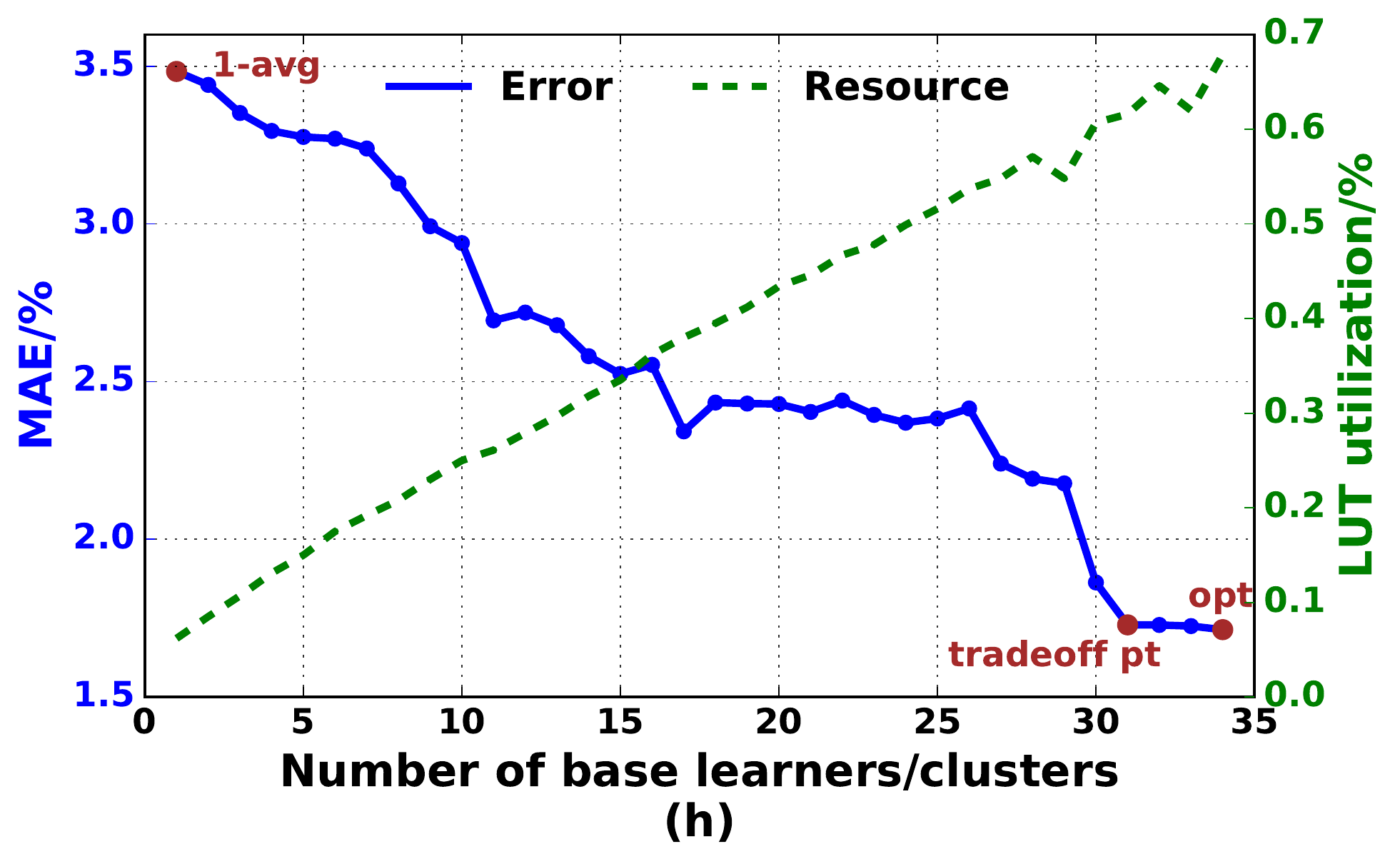}
\includegraphics[width=5.95cm]{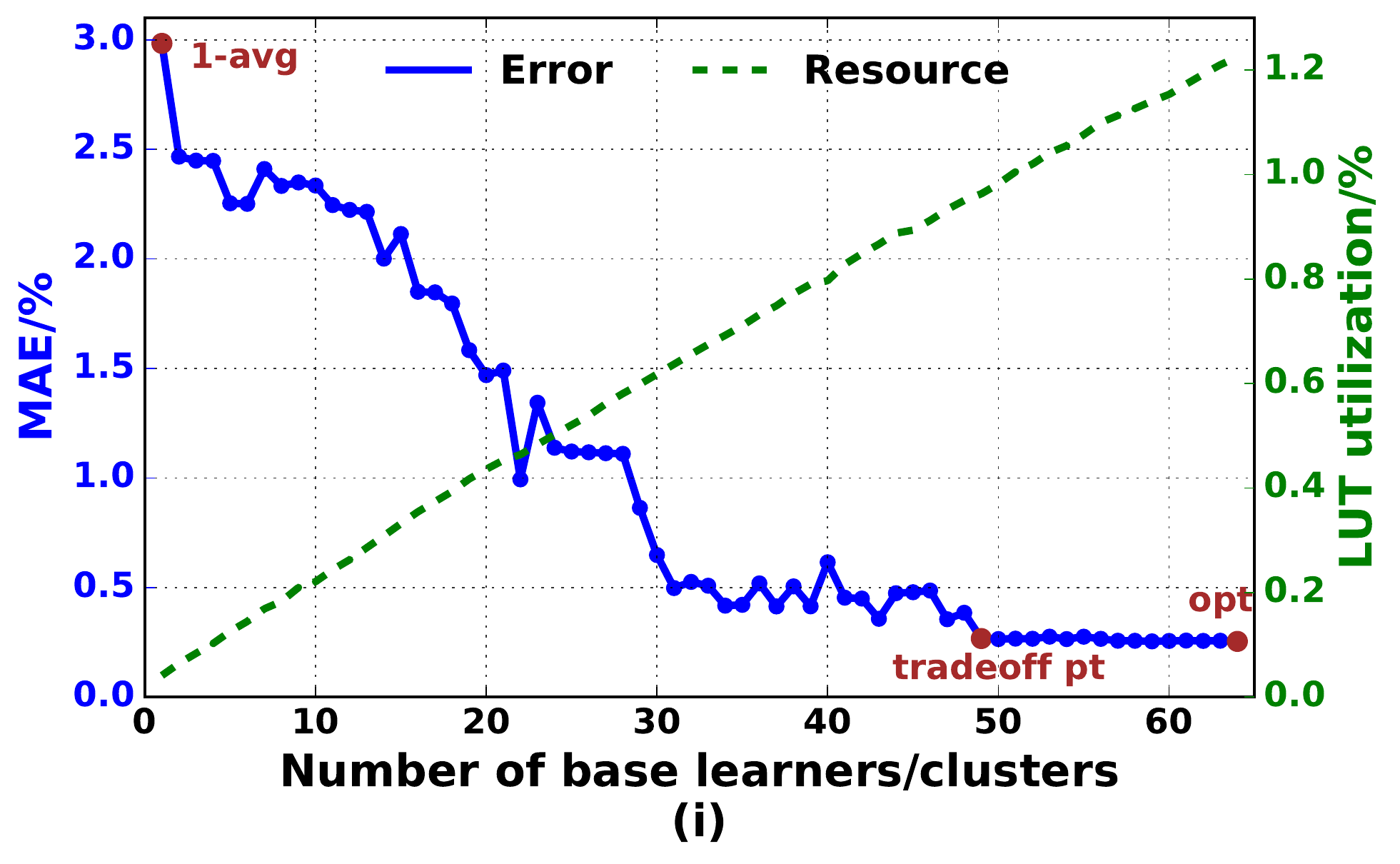}
\vspace{-5mm}
\caption{Accuracy \& LUT utilization curves of ensemble estimators deploying different numbers of base learners: (a) Atax; (b) Bicg; (c) Bbgemm; (d) Gemver; (e) Gemmncubed; (f) Matrixmult; (g) Symm; (h) Syr2k; and (i) Doitgen.}
\label{fig:ensacc}
\vspace{-6mm}
\end{center}
\end{figure*}

We study the estimation accuracy of the ensemble model using different numbers of base learners and evaluate the actual gains as the number of base learners increases. The results for different benchmarks are shown in Fig.~\ref{fig:ensacc}. It can be discovered from the curves that, in most cases, the estimation accuracy tends to increase as more base learners are deployed. This can be deduced from Equation (\ref{eq:ensvsinv}), where the error of the ensemble model, $E_{ens}$, is inversely proportional to the number of base learners, $K$. There are, however, some outliers deviating from the theoretical proof; that is, in some cases, the error goes higher when using a slightly larger number of base learners. This can be explained by the fact that the k-means algorithm cannot always guarantee that increasing the number of clusters monotonously enhances the quality of clustering. In practice, when there are some minor changes in $K$, the NP-hard k-means problem may fall into a local minimum that probably harms the clustering quality~\cite{lloyd,ost06}. Nevertheless, the existence of outliers does not alter the general trend that the error diminishes with a growing number of base learners.

In all, the accuracy of the ensemble model first decreases and then progressively stabilizes with an increase in base learners. We describe a point to optimized accuracy as the \emph{optimal point} (denoted as opt.) in every curve of Fig.~\ref{fig:ensacc}. The lowest error mostly appears when the cluster numbers are maximized. Through analyzing the ensemble results, we observe a \emph{tradeoff point} (denoted as tradeoff pt.) for a benchmark as a point in a curve offering prediction accuracy close to the optimal point, which is usually selected as the first point at which the curve converges. We elaborate the results for the accuracy at the tradeoff points and optimal points for all benchmarks in Table~\ref{table:ensacc}. We also show the estimation accuracy of the single averaging estimators (denoted as 1-avg.) as the cases of solely using one base learner.

\begin{table}[t]
\vspace{-3mm}
\caption{Accuracy of ensemble estimators at tradeoff points and optimal points, compared with the single averaging model.}
\vspace{-3mm}
\begin{center}
\begin{tabular}[width=\linewidth]{cccccc}
    \toprule
    \multicolumn{1}{c|}{\multirow{2}{*}{\textbf{Benchmark}}} & \multicolumn{2}{c|}{\multirow{1}{*}{\textbf{No. of base learners}}} & \multicolumn{3}{c}{\multirow{1}{*}{\textbf{MAE/\%}}} \\

     \multicolumn{1}{c|}{} & \multicolumn{1}{c|}{tradeoff pt.}& \multicolumn{1}{c|}{opt.}& \multicolumn{1}{c|}{1-avg.}& \multicolumn{1}{c|}{tradeoff pt.} & \multicolumn{1}{c}{opt.} \\ \midrule

   \multicolumn{1}{l|}{Atax}&
   \multicolumn{1}{c|}{21}&\multicolumn{1}{c|}{24}&\multicolumn{1}{c|}{3.82}&\multicolumn{1}{c|}{1.76}&
   \multicolumn{1}{c}{1.74} \\

   \multicolumn{1}{l|}{Bicg}&
   \multicolumn{1}{c|}{18}&\multicolumn{1}{c|}{21}&\multicolumn{1}{c|}{4.16}&\multicolumn{1}{c|}{1.90}&
   \multicolumn{1}{c}{1.87} \\

   \multicolumn{1}{l|}{Bbgemm}&
   \multicolumn{1}{c|}{22}&\multicolumn{1}{c|}{32}&\multicolumn{1}{c|}{2.72}&\multicolumn{1}{c|}{0.91}&
   \multicolumn{1}{c}{0.88} \\

   \multicolumn{1}{l|}{Gemver}&
   \multicolumn{1}{c|}{40}&\multicolumn{1}{c|}{45}&\multicolumn{1}{c|}{4.29}&\multicolumn{1}{c|}{1.50}&
   \multicolumn{1}{c}{1.47} \\

   \multicolumn{1}{l|}{Gemmncubed}&
   \multicolumn{1}{c|}{40}&\multicolumn{1}{c|}{43}&\multicolumn{1}{c|}{2.63}&\multicolumn{1}{c|}{0.71}&
   \multicolumn{1}{c}{0.69} \\

   \multicolumn{1}{l|}{Matrixmult}&
   \multicolumn{1}{c|}{53}&\multicolumn{1}{c|}{60}&\multicolumn{1}{c|}{3.74}&\multicolumn{1}{c|}{0.30}&
   \multicolumn{1}{c}{0.18} \\

   \multicolumn{1}{l|}{JPGizigzag}&
   \multicolumn{1}{c|}{43}&\multicolumn{1}{c|}{60}&\multicolumn{1}{c|}{4.75}&
   \multicolumn{1}{c|}{1.58}&\multicolumn{1}{c}{1.57} \\

   \multicolumn{1}{l|}{JPGshift}&
   \multicolumn{1}{c|}{21}&\multicolumn{1}{c|}{25}&\multicolumn{1}{c|}{4.36}&
   \multicolumn{1}{c|}{1.37}&\multicolumn{1}{c}{1.29} \\

   \multicolumn{1}{l|}{Symm}&
   \multicolumn{1}{c|}{27}&\multicolumn{1}{c|}{35}&\multicolumn{1}{c|}{4.05}&\multicolumn{1}{c|}{1.25}&
   \multicolumn{1}{c}{1.13} \\

   \multicolumn{1}{l|}{Syr2k}&
   \multicolumn{1}{c|}{31}&\multicolumn{1}{c|}{34}&\multicolumn{1}{c|}{3.48}&\multicolumn{1}{c|}{1.73}&
   \multicolumn{1}{c}{1.71} \\

   \multicolumn{1}{l|}{Doitgen}&
   \multicolumn{1}{c|}{49}&\multicolumn{1}{c|}{64}&\multicolumn{1}{c|}{2.98}&\multicolumn{1}{c|}{0.27}&
   \multicolumn{1}{c}{0.25} \\
   \bottomrule
\end{tabular}
\end{center}
\label{table:ensacc}
\vspace{-8mm}
\end{table}

The average error of the single averaging model is 3.73\%. In comparison to an average error of 3.86\% for a single decision-tree-based model, as shown in Section~\ref{subsec:dtaccuracy}, the single averaging model only exhibits an insignificant refinement. This is because there exists a high degree of correlation among error of different predictions from the same model, which counteracts the benefits of the averaging effect by introducing high correlation error, $\sum\E_{\mathbf{x}}[\epsilon_{i}(\mathbf{x})\cdot \epsilon_{j}(\mathbf{x})]$. Conversely, the average error for the ensemble estimators at the tradeoff points and at the optimal points across different benchmarks is 1.21\% and 1.16\%, respectively. It is also noteworthy that the estimators built at the tradeoff points show modest accuracy degradation in comparison to the optimal ensemble estimators, while the reduction in the number of base learners is 17\% on average. To summarize, the improvements of our proposed ensemble model at tradeoff points and optimal points over the single decision-tree-based model are 1.36--13.41$\times$ and 1.38--19.67$\times$, respectively.

\vspace{-3mm}
\subsection{Ensemble Model: Overhead}
\label{subsec:emoh}
We study the tradeoff between the accuracy and overheads of the integrated ensemble monitoring hardware, with the results shown in Table~\ref{table:ensoh}. The benchmarks with ensemble hardware showing DSP utilizations are equipped with DSP-based activity counters, while the others use LUT-based counters. Concretely speaking, we consider three factors after the integration of the ensemble hardware: i) resource usage of each base learner (bl.) along with other global control units (ctrl.), ii) the effect on operating frequencies (``+'' means increase and ``-'' means decrease) regarding each tradeoff point and optimal point, and iii) power consumption of the ensemble estimators at every one of the tradeoff points and optimal points.

Regarding the resource overhead, we discover that the LUT and RAM utilization is almost linearly related to the number of base learners, while it is relatively independent of the size of application, as depicted by the green curves in Fig.~\ref{fig:ensacc}. We can observe a tradeoff between resource overhead and accuracy in Fig.~\ref{fig:ensacc}. We compute the average resource utilization per base learner from multiple ensemble estimators with varying numbers of base learners, and deduce the resource overhead of other peripheral controllers, including three preprocessing units and an average unit, as reported in Table~\ref{table:ensoh}. The resource utilization demonstrates that each base learner can be implemented efficiently with no more than 325 LUTs and four BRAMs, which are close to the single decision tree utilization. The maximum LUT overhead of the ensemble circuit with up to 64 base learners (Doitgen) is 1.22\% of the target device. We can come to the conclusion that the proposed ensemble monitoring hardware is area-efficient and scalable. It is also worthwhile to note that~\cite{najem14} and~\cite{kapow16} induced an extra load of 5\% of the CPU time, while our proposed monitoring scheme does not require the employment of a processor.

\begin{table*}[t]
\caption{Overheads of the ensemble monitoring hardware.}
\vspace{-4mm}
\begin{center}
\begin{tabular}[width=\linewidth]{cccccccccc}
    \toprule
    \multicolumn{1}{c|}{\multirow{2}{*}{\textbf{Benchmark}}} & \multicolumn{4}{c|}{\multirow{1}{*}{\textbf{Resource} (in number)}} & \multicolumn{3}{c|}{\multirow{1}{*}{\textbf{Frequency} (MHz)}} & \multicolumn{2}{c}{\multirow{1}{*}{\textbf{Power} (mW)}}\\

   \multicolumn{1}{c|}{} & \multicolumn{1}{c|}{LUT per bl.}& \multicolumn{1}{c|}{RAM per bl.}& \multicolumn{1}{c|}{LUT for ctrl.}& \multicolumn{1}{c|}{DSP for ctrl.} & \multicolumn{1}{c|}{baseline} & \multicolumn{1}{c|}{tradeoff pt.} & \multicolumn{1}{c|}{opt.} & \multicolumn{1}{c|}{tradeoff pt.}
   & \multicolumn{1}{c}{opt.}\\ \midrule

   \multicolumn{1}{l|}{Atax}&
   \multicolumn{1}{c|}{189}&\multicolumn{1}{c|}{2.5}&\multicolumn{1}{c|}{329}&\multicolumn{1}{c|}{10}&
   \multicolumn{1}{c|}{117.65}&\multicolumn{1}{c|}{+0.15}&\multicolumn{1}{c|}{+0.42}&
   \multicolumn{1}{c|}{130}&\multicolumn{1}{c}{135} \\

   \multicolumn{1}{l|}{Bicg}&
   \multicolumn{1}{c|}{188}&\multicolumn{1}{c|}{2.5}&\multicolumn{1}{c|}{358}&\multicolumn{1}{c|}{10}&
   \multicolumn{1}{c|}{113.30}&\multicolumn{1}{c|}{-0.50}&\multicolumn{1}{c|}{+0.89}&
   \multicolumn{1}{c|}{117}&\multicolumn{1}{c}{133} \\

   \multicolumn{1}{l|}{Bbgemm}&
   \multicolumn{1}{c|}{325}&\multicolumn{1}{c|}{4}&\multicolumn{1}{c|}{246}&\multicolumn{1}{c|}{30}&
   \multicolumn{1}{c|}{105.62}&\multicolumn{1}{c|}{-0.30}&\multicolumn{1}{c|}{+0.46}&
   \multicolumn{1}{c|}{199}&\multicolumn{1}{c}{321} \\

   \multicolumn{1}{l|}{Gemver}&
   \multicolumn{1}{c|}{222}&\multicolumn{1}{c|}{3}&\multicolumn{1}{c|}{524}&\multicolumn{1}{c|}{20}&
   \multicolumn{1}{c|}{100.36}&\multicolumn{1}{c|}{-0.21}&\multicolumn{1}{c|}{+0.48}&
   \multicolumn{1}{c|}{286}&\multicolumn{1}{c}{303} \\

   \multicolumn{1}{l|}{Gemmncubed}&
   \multicolumn{1}{c|}{235}&\multicolumn{1}{c|}{3}&\multicolumn{1}{c|}{555}&\multicolumn{1}{c|}{0}&
   \multicolumn{1}{c|}{134.17}&\multicolumn{1}{c|}{-5.13}&\multicolumn{1}{c|}{-1.86}&
   \multicolumn{1}{c|}{353}&\multicolumn{1}{c}{369} \\

   \multicolumn{1}{l|}{Matrixmult}&
   \multicolumn{1}{c|}{187}&\multicolumn{1}{c|}{2.5}&\multicolumn{1}{c|}{299}&\multicolumn{1}{c|}{0}&
   \multicolumn{1}{c|}{133.74}&\multicolumn{1}{c|}{-0.85}&\multicolumn{1}{c|}{-2.66}&
   \multicolumn{1}{c|}{310}&\multicolumn{1}{c}{334} \\

   \multicolumn{1}{l|}{JPGizigzag}&
   \multicolumn{1}{c|}{194}&\multicolumn{1}{c|}{2}&\multicolumn{1}{c|}{168}&
   \multicolumn{1}{c|}{20}&
   \multicolumn{1}{c|}{102.60}&\multicolumn{1}{c|}{-2.60}&\multicolumn{1}{c|}{-2.59}&
   \multicolumn{1}{c|}{344}&\multicolumn{1}{c}{425} \\

   \multicolumn{1}{l|}{JPGshift}&
   \multicolumn{1}{c|}{280}&\multicolumn{1}{c|}{3}&\multicolumn{1}{c|}{440}&
   \multicolumn{1}{c|}{20}&
   \multicolumn{1}{c|}{125.08}&\multicolumn{1}{c|}{+0.20}&\multicolumn{1}{c|}{-0.08}&
   \multicolumn{1}{c|}{160}&\multicolumn{1}{c}{190} \\

   \multicolumn{1}{l|}{Symm}&
   \multicolumn{1}{c|}{236}&\multicolumn{1}{c|}{3}&\multicolumn{1}{c|}{557}&\multicolumn{1}{c|}{0}&
   \multicolumn{1}{c|}{125.47}&\multicolumn{1}{c|}{+0.52}&\multicolumn{1}{c|}{+0.09}&
   \multicolumn{1}{c|}{133}&\multicolumn{1}{c}{182} \\

   \multicolumn{1}{l|}{Syr2k}&
   \multicolumn{1}{c|}{230}&\multicolumn{1}{c|}{3}&\multicolumn{1}{c|}{522}&\multicolumn{1}{c|}{0}&
   \multicolumn{1}{c|}{125.02}&\multicolumn{1}{c|}{+0.69}&\multicolumn{1}{c|}{+0.90}&
   \multicolumn{1}{c|}{166}&\multicolumn{1}{c}{167} \\

   \multicolumn{1}{l|}{Doitgen}&
   \multicolumn{1}{c|}{226}&\multicolumn{1}{c|}{3}&\multicolumn{1}{c|}{279}&\multicolumn{1}{c|}{0}&
   \multicolumn{1}{c|}{120.48}&\multicolumn{1}{c|}{-5.17}&\multicolumn{1}{c|}{+0.42}&
   \multicolumn{1}{c|}{244}&\multicolumn{1}{c}{352} \\
   \bottomrule
\end{tabular}
\end{center}
\label{table:ensoh}
\vspace{-5mm}
\end{table*}

The operating frequencies for the different benchmarks before and after integrating the ensemble estimators are also shown in Table~\ref{table:ensoh}, using the original benchmarks without the monitoring hardware as the baselines. The frequencies are sometimes improved slightly after the instrument of the ensemble circuits. In all, the integration of the ensemble hardware only exerts a modest impact on the performance, with a maximum degradation of around 5 MHz (4.3\%) on the operating frequencies.

With respect to the power overhead, it is reasonable to conceive that the power consumption increases with the number of base learners, which also reveals a tradeoff between the power overhead and power estimation accuracy. The power evaluation results in Table.~\ref{table:ensoh} demonstrate that the power overhead is within 425 mW for the ensemble hardware. To abate power overhead, we can diminish the number of base learners at the cost of higher error. With this objective, using the tradeoff point instead of the optimal point to build an ensemble model can reduce the power overhead by a maximum of 122 mW, with an insignificant deterioration in accuracy.

\section{Conclusion}
\label{sec:conc}
Power consumption never fails to be an important design consideration for FPGAs. In this work, we establish a dynamic power monitoring scheme within the timescale of hundreds of execution cycles to facilitate emerging fine-grained power management strategies. We introduce a novel and specialized ensemble model for runtime dynamic power monitoring in FPGAs. In the model, every decomposed base learner is established using a non-overlapping data set derived from a specific cluster of FSM states with homogeneous activity patterns. To assist in this goal, we describe a generic and complete CAD flow from sample generation to feature selection, FSM state clustering, hyperparameter tuning and model ensemble. In order to put the ensemble model into practice for real-time power monitoring, we propose an efficient hardware realization, which can be embedded into the target application.

In the experiments, we first present the results of a single decision-tree-based power model, which offers a 2.41--6.07$\times$ improvement in prediction accuracy in comparison to the traditional linear model. Furthermore, we study the extra gains in prediction accuracy derived from our customized ensemble model, by varying the number of base learners. We observe a tradeoff between estimation accuracy and overheads of resources, performance and power introduced by the ensemble monitoring hardware. Through analyzing the experimental results, we find a tradeoff point for accuracy and overheads, and an optimal point for accuracy, which, respectively, show a capability to achieve error 1.36--13.41$\times$ and 1.38--19.67$\times$ lower than the single decision-tree-based model. The hardware implementation of runtime monitoring with up to 64 base learners incurs modest overheads: 1.22\% of LUT utilization, 4.3\% of performance and less than 425 mW power dissipation. By using the tradeoff points instead of optimal points to construct the monitoring hardware, the number of base learners can be reduced by 17\% on average, and the power consumption can be accordingly cut by up to 122 mW, at the cost of an insignificant accuracy loss.

In summary, the proposed ensemble model provides accurate dynamic power estimate within an error margin of 1.90\% of a commercial gate-level power estimation tool. The hardware realization of runtime power monitoring demonstrates low overheads of resource, performance and power consumption.

\section*{Acknowledgment}
This work is funded by Hong Kong RGC GRF 16245116.

\bibliographystyle{IEEEtran}
\bibliography{Ref}

\begin{IEEEbiography}[{\includegraphics[width=1in,height=1.25in,clip,keepaspectratio]{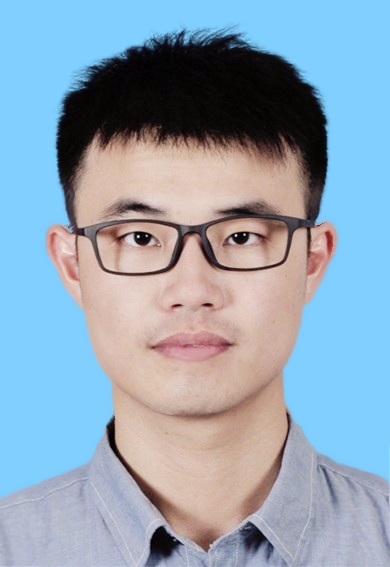}}]{Zhe Lin}
(S'15) received his B.S. degree from School of Electronic Science and Engineering from Southeast University, Nanjing, China, in 2014. Since 2014, he has been a Ph.D. Student in the Department of Electronic and Computer Engineering at Hong Kong University of Science and Technology (HKUST), Hong Kong. Zhe's current research interests cover FPGA-based heterogeneous multicore systems and power management strategies of modern FPGAs.
\end{IEEEbiography}

\begin{IEEEbiography}[{\includegraphics[width=1in,height=1.25in,clip,keepaspectratio]{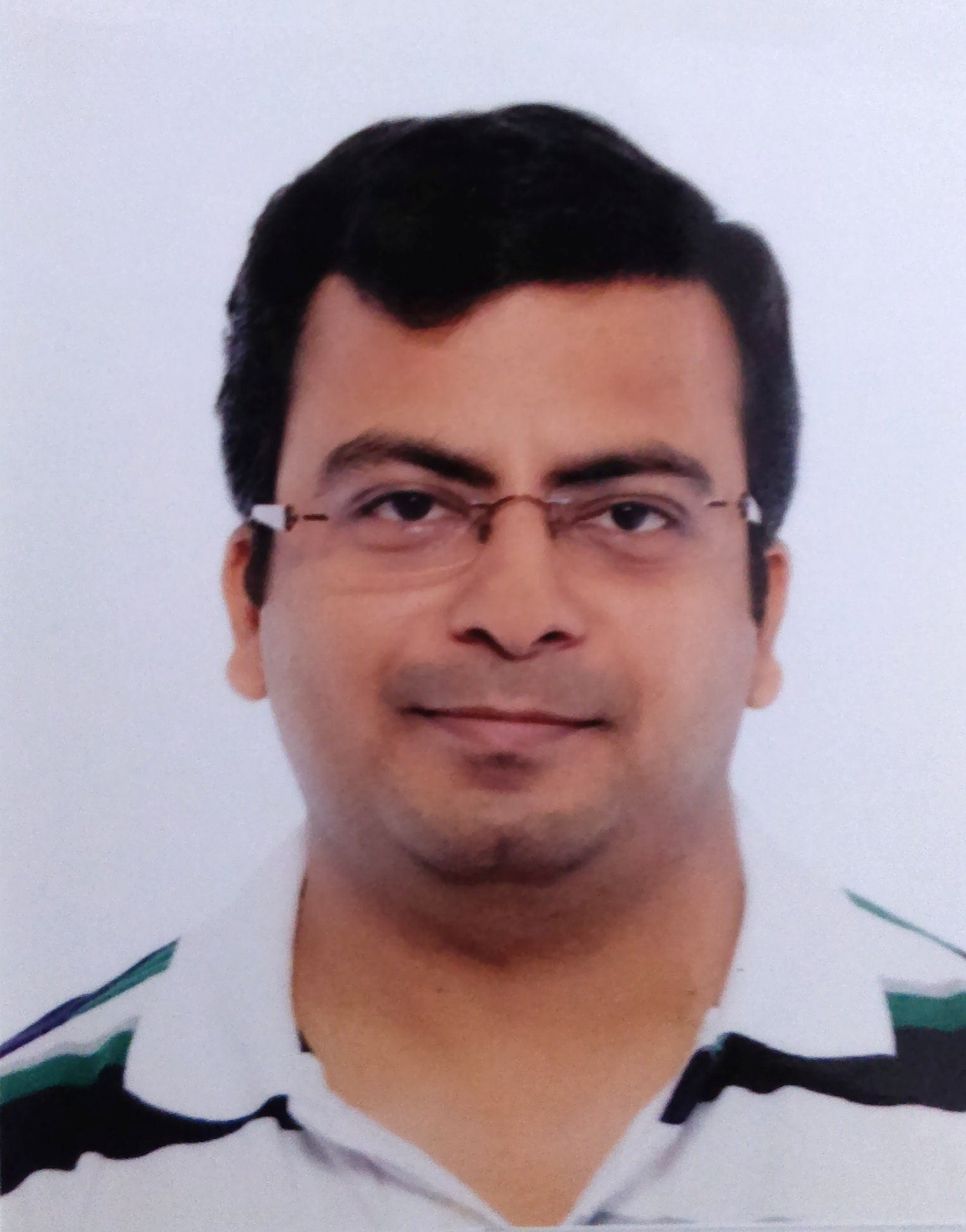}}]{Sharad Sinha}
(S'03, M'14) is an assistant professor with Dept. of Computer Science and Engineering, Indian Institute of Technology (IIT) Goa. Previously, he was a Research Scientist at NTU, Singapore. He received his PhD degree in Computer Engineering from NTU, Singapore (2014). He received the \textit{Best Speaker Award} from \textit{IEEE CASS Society}, Singapore Chapter, in 2013 for his PhD work on High Level Synthesis and serves as an Associate Editor for \textit{IEEE Potentials} and \textit{ACM Ubiquity}. Dr. Sinha earned a Bachelor of Technology (B.Tech) degree in Electronics and Communication Engineering from Cochin University of Science and Technology (CUSAT), India in 2007. From 2007-2009, he was a design engineer with Processor Systems (India) Pvt. Ltd. Dr. Sinha's research and teaching interests are in computer arhcitecture, embedded systems and reconfigurable computing.
\end{IEEEbiography}

\begin{IEEEbiography}[{\includegraphics[width=1in,height=1.25in,clip,keepaspectratio]{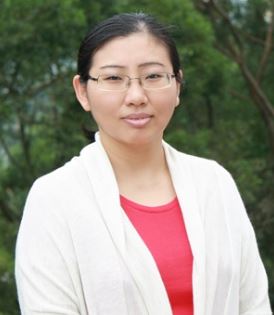}}]{Wei Zhang}
(M'05) received a Ph.D. degree from Princeton University, Princeton, NJ, USA, in 2009. She was an assistant professor with the School of Computer Engineering, Nanyang Technological University, Singapore, from 2010 to 2013. Dr. Zhang joined the Hong Kong University of Science and Technology, Hong Kong, in 2013, where she is currently an associated professor and she established the reconfigurable computing system laboratory (RCSL).

Dr. Zhang has authored or co-authored over 80 book chapters and papers in peer reviewed journals and international conferences. Dr. Zhang serves as the Associate Editor for \textit{ACM Transactions on Embedded Computing Systems (TECS)}, \textit{IEEE Transactions on Very Large Scale Integration Systems (TVLSI)}, and \textit{ACM Journal on Emerging Technologies in Computing Systems (JETC)}. She also serves on many organization committees and technical program committees. Dr. Zhang’s current research interests include reconfigurable systems, FPGA-based design, low-power high-performance multicore systems, electronic design automation, embedded systems, and emerging technologies.
\end{IEEEbiography}

\end{document}